\documentclass[pdflatex,sn-vancouver-num]{sn-jnl}
\usepackage{graphicx}
\usepackage{lipsum}
\usepackage{rotating}
\usepackage[normalem]{ulem}
\usepackage{tikz}
\usepackage{pgfplots}
\usepackage{booktabs}
\usepackage{graphicx}
\usepackage{hyperref}
\hypersetup{hypertexnames=false}
\usepackage{amsmath}
\usepackage{amssymb}
\usepackage{array}
\usepackage{algorithm}
\usepackage{algpseudocode}
\usetikzlibrary{arrows.meta,positioning,shapes.geometric,calc,shadows}
\pgfplotsset{compat=1.18}
\definecolor{rsxteal}{RGB}{0,77,64}
\definecolor{rsxsky}{RGB}{30,136,229}
\definecolor{rsxlight}{RGB}{230,242,255}
\definecolor{rsxgreen}{RGB}{0,128,0}
\definecolor{rsxorange}{RGB}{230,126,34}
\definecolor{rsxred}{RGB}{192,57,43}
\definecolor{rsxmagenta}{RGB}{216,27,96}
\definecolor{rsxcoral}{RGB}{255,101,93}
\definecolor{rsxsunshine}{RGB}{241,219,75}
\author*[1,2]{\fnm{Rohit} \sur{Goswami}}\email{rgoswami@ieee.org}
\author[3]{\fnm{Ruhila} \sur{Goswami}}\email{rug17@hi.is}
\affil[1]{\orgname{TurtleTech ehf.}, \orgaddress{\city{Reykjav\'{\i}k}, \country{Iceland}}}
\affil[2]{\orgdiv{Institute IMX and Laboratory of Computational Science and Modelling (Lab-COSMO)}, \orgname{\'{E}cole Polytechnique F\'{e}d\'{e}rale de Lausanne (EPFL)}, \orgaddress{\street{Station 12}, \city{Lausanne}, \postcode{1015}, \country{Switzerland}}}
\affil[3]{\orgdiv{Faculty of Life and Environmental Sciences}, \orgname{University of Iceland}, \orgaddress{\city{Reykjav\'{\i}k}, \country{Iceland}}}
\date{2026}
\title{rsx: a high-performance streaming toolkit for RAD-seq sex determination}
\hypersetup{
 pdfauthor={Rohit Goswami, Ruhila Goswami},
 pdftitle={rsx: a high-performance streaming toolkit for RAD-seq sex determination},
 pdfkeywords={},
 pdfsubject={},
 pdfcreator={Emacs 30.2 (Org mode 9.7.11)}, 
 pdflang={English}}
\begin{document}

\abstract{
\par\noindent\textbf{Background.} Restriction site-associated DNA sequencing (RAD-seq) is widely used to discover sex-linked markers in non-model organisms, and RADSex provides the reference workflow for building marker-by-individual depth tables and testing sex-biased marker distributions. Its table-building commands grow memory-hungry as panels reach millions of RAD tags, it reports frequentist calls with no posterior evidence, and it offers no Python or C interface.
\par\noindent\textbf{Results.} rsx is a Rust implementation of the complete RADSex command set that preserves marker-table semantics and command-line compatibility. It combines 2-bit DNA keys, parallel ingestion, memory-mapped tables, external sorting, bitset group counts and a streamed Gram matrix so that writable allocations stay bounded by the number of individuals or by an explicit buffer, with false-discovery-rate ranking the one deliberate exception. Conjugate Beta-Binomial Bayes factors and directional posteriors grade each marker as a strict call, a posterior-supported hypothesis or a Bayes-factor-only row, and an optional CUDA backend batches the per-marker arithmetic on the GPU. On four published RAD-seq panels comprising 41.9 billion sequenced bases, rsx reproduced the RADSex v1.2.0 calls, recovered every Bonferroni-significant positive-control marker, and was 8.38-fold faster in geometric mean across 56 paired timings; the CUDA backend adds up to 29.86-fold on the p-value batch. Python and C bindings drive the same core from notebooks and pipelines.
\par\noindent\textbf{Conclusions.} rsx is an allocation-bounded, statistically extended replacement for RADSex that stays backward-compatible and reports its evidence in explicit grades. It is released under the GPL-3.0-or-later licence, with a reproducibility archive covering every reported number.
}


\maketitle
\section{Background}
\label{sec:orga79b57c}

\subsection{RAD-seq and sex determination}
\label{sec:orga75c6d2}

Restriction site-associated DNA sequencing (RAD-seq) samples short
genomic regions next to restriction sites. The original RAD marker work
showed that these tags support high-density mapping without whole-genome
sequencing of every individual \cite{miller2007rad,baird2008rad}. Later ecological and
evolutionary genomics work established RAD-seq as a practical design for
non-model systems where many individuals outweigh complete genomes
\cite{davey2011radseq,davey2011markerdiscovery,andrews2016radseq,hohenlohe2010stickleback}.
The same methodological family includes double-digest RAD designs for
cost-effective de novo SNP discovery \cite{peterson2012ddrad}, genotyping-by-sequencing \cite{elshire2011gbs}, and
well-known genotyping pipelines such as Stacks, whose later paired-end
extensions illustrate how much of the field depends on careful treatment
of short-read RAD loci \cite{catchen2011stacks,rochette2019stacks2}.
RAD-seq data carry statistical power alongside systematic artifacts. Allele
dropout, restriction-site variation, and depth heterogeneity each demand
explicit treatment before any RAD tag counts as biological evidence
\cite{davey2013radseqfeatures}.

Sex-determination studies apply the same reduced-representation principle
to a defined inference task: locate markers whose presence, absence, or
depth tracks phenotypic sex. A marker seen mostly in
males is a putative Y-linked marker under XX/XY inheritance; a marker
seen mostly in females is a putative W-linked marker under ZZ/ZW
inheritance \cite{bachtrog2014sex}. rsx tests whether the counts fit sex-limited
inheritance. Gamble
and Zarkower \cite{gamble2014radsexmarkers} defined the marker-discovery
logic for sex-specific RAD tags; Gamble \cite{gamble2016recognize}
showed that such markers can reveal sex chromosome systems when
cytogenetic data provide no clear signal; such systems range from young,
homomorphic sex chromosomes to old, degenerate ones \cite{bachtrog2013y}, and
in teleost fish they turn over especially fast \cite{kottler2018teleost}. Later studies align the same tags to
draft genomes to test physical clustering \cite{fowler2016rockfish}.

The \emph{P. altivelis} RADSex positive-control result
matches an ayu XX/XY system: prior work detected ayu sex-linked markers
\cite{watanabe2004ayu}, sequencing found male-specific markers consistent
with XX/XY \cite{li2021ayu}, and later work identified a Y-linked
amhr2bY sex-determining gene \cite{nakamoto2021ayu}. By contrast, RAD panels
resist interpretation for species with weak, polygenic, labile, or
population-specific sex association. Zebrafish RAD mapping shows
sex-associated regions that vary across populations and lack a shared
chromosome-wide signal \cite{anderson2012zebrafish,wilson2014wild}.
rsx reports sex-linked marker calls and hypotheses with their evidence
grades.

RADSex \cite{feron2021radsex} provides the matched reference implementation:
it accepts the same RAD-seq reads and marker tables and implements the same
marker-level commands. We therefore use RADSex for controlled runtime and
output comparisons. SEX-DETector and findZX answer related biological
questions from different assays and inferential units. Whole-genome methods
provide another distinct class of inputs. The
method-scope comparison and executed method-native examples appear in Appendix
A.12.
\subsection{Limitations of the C++ implementation}
\label{sec:org89b37bf}

The original C++ RADSex remains the reference for the workflow. rsx
addresses three practical limits that surface once RAD-seq marker tables
grow large: commands that accumulate records grow sensitive to table
size; independent tables require an out-of-core merge; and a single
chi-squared/Bonferroni path leaves no statistical alternatives.
\subsection{Novelty of rsx}
\label{sec:org07fa32d}

rsx extends RADSex with streaming algorithms, exact-count compatibility,
and explicit posterior evidence. The biological question is unchanged,
whether a RAD tag's presence tracks phenotypic sex, but the algorithms
underneath now run in bounded memory over the full table, and a Bayesian
layer turns the same counts into a probability of sex linkage. rsx keeps
the marker-table semantics that make results comparable with published
RADSex studies, and adds posterior sex-linkage probabilities, full-table
bounded-memory execution, and auditable numerical kernels
(Table \ref{tab:novelty}).

\begin{table}[h]
\centering
\begin{tabular}{p{0.28\linewidth}p{0.62\linewidth}}
\toprule
Core contribution & What it adds for inference \\
\midrule
Compatible marker-table semantics & reproduces published RADSex calls on the same biological datasets, so new outputs can be interpreted relative to prior sex-determination results \\
Redone full-table algorithms & memory-mapped parsing, bitset group counts, external sort/merge, and two-pass thresholding avoid downsampling rare candidate markers \\
Posterior sex-linkage calls & adds Bayes factors and posterior P(sex-linked) for the same marker counts, separating putative Y-linked or W-linked markers from Bayes-factor-only rows \\
Numerical provenance & records derivations for erfc p-values, sparse medians, and streaming Gram PCA so the optimized kernels are mathematically checkable \\
Workflow bindings & exposes the same core through Python and C interfaces so biological analyses can continue in notebooks, R/Python workflows, and downstream visualisation code \\
\bottomrule
\end{tabular}
\caption{Algorithmic and inferential novelty of rsx relative to the
original RADSex workflow.}
\label{tab:novelty}
\end{table}

rsx works at the level of the marker table that RADSex users already
produce, and from that one input it gives exact same-command
compatibility, faster full-table execution, ranked sex-linked marker
evidence below the strict Bonferroni threshold, and quality-control (QC)
outputs that flag when a signal is driven by sparse marker presence. Where
whole-genome or transcriptome data are available, scaffold- and gene-level
methods resolve sex linkage directly; rsx extracts more evidence, and more
auditable evidence, from the reduced-representation data that most
non-model studies actually have.
\section{Implementation}
\label{sec:org1cf9d0b}

\subsection{Architecture}
\label{sec:org2dd90d1}

The Rust workspace contains three crates:
\begin{itemize}
\item \texttt{rsxcore}: library crate with all algorithms; it produces an \texttt{rlib} plus
static and dynamic C libraries (\texttt{staticlib} and \texttt{cdylib})
\item \texttt{rsx-cli}: command-line interface via clap, installed as the \texttt{rsx} binary
\item \texttt{pyrsx}: Python bindings and a Click-based Python CLI
\end{itemize}

Rust-Bio \cite{koester2016rustbio} established the precedent for Rust in
bioinformatics tooling. rsx shares its goal, memory safety for
performance-critical sequence analysis, but specializes the core around
RADSex-compatible marker-table semantics. Exposing one validated core
through a command-line interface and two bindings (Python and C) uses a narrow C ABI with
thin language bindings layered above it. The
pattern is long-established, running through decades of Fortran-to-C and
Fortran-to-C++ interoperability
\cite{gray1999shadow,decyk2008passing,pletzer2008fortran}, and it still
underlies current scientific interface tooling such as f2py
\cite{stateOfFortran2022}. The algorithms live in one crate and the
bindings call that shared core. Tests exercise all three language entry
points against the same fixtures; the notebook and command-line
paths therefore return the same result. We used the same structure in an
earlier scientific library \cite{Goswami2020}.

Beyond the default build, Cargo features control optional capabilities. The
\texttt{merge} command can write Parquet with ZSTD compression for downstream
columnar tooling, and \texttt{process} ingestion can run across nodes under MPI. A
separate feature controls the minimap2-backed \texttt{map} command. The other commands
can build where minimap2 is unavailable (for example on Windows).

The repository applies \texttt{rustfmt} formatting, \texttt{clippy} static analysis, unit and
integration tests, cross-language binding tests, and executable SymPy and Lean
checks for the numerical identities. The authors used large-language-model-based
coding assistants while writing the rsx source code, mainly for routine
implementations and test scaffolding; every change passed author review and the
validation reported under Software validation. No such tool wrote or edited this
manuscript.

The C-compatible ABI follows the metatensor \cite{metatensor} pattern of
status codes, thread-local errors, and \texttt{catch\_unwind} panic safety for
the exported functions. cbindgen generates headers from
\texttt{\#[repr(C)]} types for downstream binding work.

Every command can be represented by a schema-versioned TOML profile. Explicit
command-line values override the input profile, after which rsx writes a
hydrated TOML containing every resolved value before validating inputs or
starting the calculation. The Python package validates this profile schema
with strict Pydantic v2 models. A discriminated union keyed by \texttt{family}
requires \texttt{probability} for a fixed prevalence or positive \texttt{alpha} and \texttt{beta}
shapes for a Beta prevalence, and unknown keys are rejected. The checked JSON
Schema covers every command profile as well as the shared Bayesian model.
Every command-line and binding surface exposes the prior parameters. With
\texttt{-{}-{}reproducibility-archive <path>}, rsx atomically writes a software-only ZIP
before the run begins. It contains the input and hydrated profiles, version and
commit metadata, build manifest, dependency SBOM, schema, \texttt{CITATION.cff},
license, and checksums, but not potentially large result files. The archive and
hydrated profile therefore remain available when validation or computation
fails, and a later software release cannot reinterpret an old input without
warning.
\subsection{Bounded-memory streaming}
\label{sec:orgc6f7f52}

The commands never materialize the full marker table unless the chosen
method requires global ranking. Table-oriented commands stream over a
memory-mapped file; merge and depth use external sorts; PCA accumulates a
Gram matrix over individuals. One procedure cannot work this way. Controlling
the false-discovery rate by the Benjamini-Hochberg method
\cite{benjamini1995controlling}, described with the other inference procedures
under Statistical methods, ranks every marker against every other, so it needs
all the p-values and their rows at once.

For a marker-table command, rsx opens the tab-separated file read-only and
maps its bytes into the process address space with \texttt{memmap2}. The parser
finds line and field boundaries directly in that byte slice and reuses one
marker buffer; it does not copy the file into a second input buffer or
retain parsed rows. The mapping reserves virtual address space, while the
operating system loads pages on demand and may evict clean pages under
memory pressure. The Bonferroni scan described below walks the table twice and
usually finds the second pass already in the page cache, which saves reading it
again but changes no result. Linux memory overcommit
governs reservations for anonymous allocations. The read-only file mapping
reserves the file's virtual address range independently. Physical residency remains subject to the kernel's
page-cache policy. Output bitsets and buffers determine the process's writable
memory demand, including the selected external-sort buffer. Linux maximum RSS
counts resident file-backed pages as well as private writable pages. A complete
scan can therefore report an RSS near the input-table size even though the
parser has not copied or retained the table. We report that process-level
quantity directly and distinguish it from the bounded writable allocation.

Ingestion runs in parallel. Many threads fold their reads into one shared,
concurrent sharded count table keyed by the packed 2-bit sequence, so the
same tag seen on different threads enters the same bucket. rsx sizes that
table from the input file size and caps the \emph{initial allocation} at four
million entries. A 64-byte-per-marker planning estimate makes this an
initial reservation of about 256 MB before allocator and table overhead.
The value does not cap the number of markers retained: the table grows when
more than four million distinct tags occur, so the setting cannot drop a
marker or create a false negative. The lower bound of 1,024 entries avoids
small repeated allocations. Each key stays in its packed 2-bit form, held
inline for the common short tag lengths, so the counting loop needs no extra
heap allocation per unique tag. The parallel count is identical to the serial
one, only faster.

\begin{table}[h]
\centering
\begin{tabular}{lll}
\toprule
Command & Memory & Strategy \\
\midrule
distrib, freq & $O(n_{\text{ind}})$ & Streaming accumulator \\
signif, subset & $O(n_{\text{ind}})$ & Two-pass mmap for Bonferroni/none \\
map & $O(\text{genome index})$ & Two-pass + minimap2 \\
depth & $O(\text{buffer})$ & Sparse external sort \\
merge & $O(\text{buffer})$ & Chunked external sort \\
pca & $O(n_{\text{ind}}^2)$ & Streaming Gram matrix \\
\bottomrule
\end{tabular}
\caption{Memory complexity of rsx commands for streaming modes.
$n_{\text{ind}}$ = number of individuals, typically 10--200. Users choose the
buffer size for external-sort commands. FDR correction materializes
p-values and row data for ranking.}
\end{table}
\subsection{Numerical algorithms}
\label{sec:orgfa56bda}

Every memory-saving kernel in rsx carries an explicit invariant: it computes
the same quantity as the textbook algorithm with less memory, or it uses an
approximation whose error we measure. Measure, not quote: for the chi-squared
tail we report what each candidate evaluator does against the reference, not
what its generator advertises. The optimized code is then auditable, and a
marker call changes when the statistical model changes rather than because a
large table forced a silent shortcut. Full forward-error bounds, the Cg formatter validation, and the unit-test
coverage appear in Appendix A.
\subsubsection{Chi-squared p-value via erfc identity}
\label{sec:org85d8fe0}

Each marker induces a \(2 \times 2\) table: present and absent counts in
the two sex groups. The Pearson \cite{pearson1900chi} or Yates-corrected \cite{yates1934contingency} statistic reduces to a single
non-negative number \(x\). For one degree of freedom,
\(\chi^2_1\) has the same distribution as \(Z^2\) for a standard normal
random variable \(Z\). Therefore, the upper-tail probability is
\begin{equation}
  P(\chi^2_1 \ge x)
  = P(|Z| \ge \sqrt{x})
  = \operatorname{erfc}\left(\sqrt{x/2}\right).
\end{equation}

A general chi-squared cumulative distribution thus becomes one call to
\(\operatorname{erfc}\), and that call is the only thing rsx changes:
\begin{equation}
  p = \operatorname{erfc}\left(\sqrt{x/2}\right).
\end{equation}
rsx does not change the statistical test, the null distribution, or the
Bonferroni threshold. When the degree of freedom is known, the implementation avoids a general
regularized-gamma call, giving a shorter and more portable numerical path. A symbolic derivation script
verifies the identity, and numerical checks reproduce standard
chi-squared \cite{pearson1900chi} critical values. Full forward-error
analysis and performance numbers appear in Appendix A.6.
\subsubsection{Platform-independent erfc evaluation}
\label{sec:org12341f9}

The reported results evaluate the identity through the platform math library,
and the CUDA kernels call the device erfc. rsx also carries a table-driven
evaluator for hosts where the platform function is unavailable or is not
trusted to give the same answer as another host, so that a p-value near a
threshold does not depend on which machine produced it. It covers \([0,6)\) with
24 panels of width \(1/4\), each a degree-14 minimax fit in the centred variable
\(u=t-c\); beyond \(t=6\) it returns zero, where
\(\operatorname{erfc}(6)\approx2.15\times10^{-17}\) already lies under the
\(10^{-16}\) floor at which rsx reports p-values.

The error we quote is the one we measured against the platform function, which
is not the one the fitting tool reports. Sollya bounds the fitted polynomial at
\(1.1\times10^{-16}\) relative; evaluating that polynomial in binary64 costs a few
units in the last place more, and the measured figure is \(4.4\times10^{-16}\)
for \(t<4.75\), covering every region where a Bonferroni or FDR threshold falls
on these panels. It reaches \(2.8\times10^{-15}\) by \(t=6\), where the p-value is
below the reporting floor anyway.

The two figures stay within a factor of four because the panels are narrow and
the variable is centred, so no monomial the evaluation touches grows beyond
order one. Fit the same function in one piece across the same interval in the
raw variable and the monomials reach \(6^{40}\), at which point the fitted bound
says nothing about the result. Both evaluators cost the same on the benchmark
host, so the table is worth having for platform independence, not for speed.
\subsubsection{CUDA marker-evidence backend}
\label{sec:orga841938}

The optional \texttt{cuda} build feature adds a GPU backend to \texttt{signif}, \texttt{triage}, and
\texttt{pca}. Users select it with \texttt{-{}-{}backend cuda}; \texttt{-{}-{}backend cpu} remains the
default. The device path covers each per-marker association test the CPU
offers, the Bayes factor with its directional posterior, and the Gram
matrix that sample PCA accumulates. Requesting the backend without a
working device returns an error rather than changing the requested calculation
or falling back without notice.

The kernels are compiled at run time for the compute capability the detected
device reports, because the Gram accumulation combines partial sums with
double-precision atomic addition.

Two machines carry every measurement reported here. One is a cluster node with
an NVIDIA A100-PCIE-40GB accelerator and dual Intel Xeon Gold 6248R processors,
96 logical CPUs and 193 GiB of memory, running under Slurm; we call it Elja
after the cluster it belongs to. The other is a workstation with an NVIDIA
GeForce RTX 5070 and an AMD Ryzen 9 9950X, 32 logical CPUs and 46 GiB of
memory, which we call Terra. The pair spans a data centre accelerator with
full-rate binary64 arithmetic and a consumer card without it.

Each marker contributes two 32-bit presence counts. The backend copies these
eight-byte pairs directly to device memory and launches a CUDA C kernel that
evaluates the same binary64 Yates statistic and erfc tail probability as the
CPU implementation. Each marker returns one binary64 p-value, so both transfer
directions move eight bytes per marker. Device setup, runtime compilation, and
the loaded kernel are retained for the process lifetime. The largest allocated
device buffers and page-locked host result buffer are also retained for later
batches. This design separates the one-time setup cost from repeated batch
execution and avoids an intermediate host count array.

The backend records kernel time and both transfer times. It also records setup,
transferred bytes, and whether the page-locked result allocation was reused.
Tests compare every GPU p-value with the scalar CPU result and compare
complete \texttt{signif} output files. These tests run on a CUDA device; selecting the
GPU backend without a working device is an error.

The accelerated work is the per-marker arithmetic and the PCA Gram
accumulation.
Marker-table mapping and parsing, group-mask population counts, multiple-test
correction, output filtering, and file writing remain on the CPU. Reported
batch speedups include both transfers and device synchronization but exclude
those surrounding command stages. The complete-command measurements below
therefore remain the appropriate guide for short, one-shot analyses.

Table \ref{tab:cuda_coverage} reports the device path for every accelerated
kernel on both hosts. Fisher's exact test gains most because it walks the
hypergeometric tail once per individual for every marker, which makes it the
one test whose cost is dominated by double-precision transcendentals; the A100
carries full-rate binary64 arithmetic where the consumer card does not, and the
two devices separate by an order of magnitude on that kernel alone. The Gram
accumulation returns identical totals because depth products are integers held
exactly in binary64, so combining partial sums in any order gives the same
result.

\begin{table}[h]
\centering
\begin{tabular}{lrrrr}
\toprule
Accelerated kernel & Markers & A100 & RTX 5070 & Agreement \\
\midrule
Yates chi-squared & 50,000,000 & 25.9x & 15.1x & $2.7\times10^{-16}$ \\
Fisher exact & 50,000,000 & 486.7x & 20.0x & $2.8\times10^{-14}$ \\
G-test & 50,000,000 & 102.7x & 24.7x & $1.2\times10^{-15}$ \\
Beta--Binomial Bayes factor & 50,000,000 & 70.7x & 14.9x & $5.7\times10^{-14}$ \\
Directional posterior & 50,000,000 & 70.7x & 14.9x & $7.2\times10^{-15}$ \\
Sample-PCA Gram & 2,000,000 & 20.6x & 5.4x & exact \\
\bottomrule
\end{tabular}
\caption{Device total speedup for every accelerated kernel at the largest
measured batch, as the median of five repetitions. Total speedup covers both
transfer directions and synchronization, so these ratios are comparable with
the last column of Table \ref{tab:cuda_elja} rather than with its kernel
column. Agreement is the worst relative
difference from the scalar path across the whole batch. The devices are an
NVIDIA A100-PCIE-40GB (the Elja node) and an NVIDIA GeForce RTX 5070 (the Terra
workstation), both with 48 individuals. Every median here derives from the per-repetition rows deposited
as \texttt{cuda\_coverage\_elja\_a100.csv} and
\texttt{cuda\_coverage\_terra\_rtx5070.csv}, so each ratio can be
recomputed from the raw measurements.}
\label{tab:cuda_coverage}
\end{table}
\subsubsection{Sparse median}
\label{sec:orgd0634e5}

Depth QC reports per-sample read-depth summaries across all retained
markers. The depth vector for one individual is mostly zeros. Let \(n\) be the number of marker depths,
\(z\) the number of zeros, and let
\(a_0 \le a_1 \le \cdots \le a_{n-z-1}\) be the sorted non-zero depths.
For a zero-based rank \(r\), define the sparse lookup
\begin{equation}
  q(r) =
  \begin{cases}
    0, & r < z,\\
    a_{r-z}, & r \ge z.
  \end{cases}
\end{equation}
Writing \(\ell=\lfloor(n-1)/2\rfloor\) and \(u=\lfloor n/2\rfloor\), rsx reports
the mathematical median
\begin{equation}
  \operatorname{median} = \frac{q(\ell)+q(u)}{2}.
\end{equation}
The ranks coincide for odd \(n\); for even \(n\) they select the adjacent middle
depths, so half-integer medians are preserved. The rank formula equals the
median of the fully sorted vector. The external sort stores only the non-zero
depths; a single count represents the zeros. The depth summary therefore
matches sorting the full vector, while the I/O volume falls by roughly the
inverse sparsity.
\subsubsection{Sample PCA via a streamed Gram matrix}
\label{sec:org0d598d0}

rsx includes PCA as a sample-level QC view. PCA does not identify
sex-linked markers on its own; it asks whether the full marker-depth
profile separates samples by sex, library size, batch, or outlying
individuals. Let \(X \in \mathbb{R}^{m \times n}\) be the marker-depth
matrix, with \(m\) markers and \(n\) individuals. In RAD-seq applications
\(m \gg n\): millions of markers but tens to hundreds of individuals.

The sample PCA depends on the centered sample Gram matrix
\begin{equation}
  C = (X-\mathbf{1}\mu^T)^T (X-\mathbf{1}\mu^T),
\end{equation}
where \(\mu\) is the vector of per-individual mean depths over markers.
Expanding the product gives the streaming formula
\begin{equation}
  C = X^T X - m \mu \mu^T.
\end{equation}
Thus, rsx streams one marker row \(x_i\) at a time, accumulates
\(X^T X \leftarrow X^T X + x_i^T x_i\) and the column sums needed for
\(\mu\), then eigendecomposes only the \(n \times n\) matrix \(C\).

If \(\tilde{X}=X-\mathbf{1}\mu^T\) and
\(\tilde{X}=U\Sigma V^T\) is the singular value decomposition, then
\begin{equation}
  C = \tilde{X}^T \tilde{X} = V \Sigma^2 V^T.
\end{equation}
The eigenvectors of \(C\) therefore equal the right singular vectors
of the centered marker table, the same per-sample PCA loadings the full
matrix would yield. The memory cost drops from \(O(mn)\) to \(O(n^2)\),
about 320 KB for 200 individuals. The implementation records
eigenvalues, variance fractions, and per-individual loadings so the
reader can tell whether a component tracks phenotypic sex or a technical
depth effect.
\subsection{Statistical methods}
\label{sec:org0c5907e}

rsx adds Fisher's exact test \cite{fisher1922interpretation} (exact for small counts), the G-test
\cite{wilks1938large} (log-likelihood ratio), Benjamini-Hochberg \cite{benjamini1995controlling} FDR control, and a conjugate
Bayesian layer to the original Yates-corrected \cite{yates1934contingency} chi-squared plus
Bonferroni \cite{bonferroni1936teoria} path (see Clark et al. \cite{clark2019polyrad} for a related
Beta-Binomial treatment of RAD-seq data on sex chromosomes). Bonferroni bounds
the family-wise error rate, the probability of even one false sex-linked
call across the whole experiment, and is correspondingly conservative.
Benjamini-Hochberg instead bounds the false-discovery rate \cite{storey2003genomewide}, the expected
fraction of false calls among those reported, which recovers power when a
panel contains many true signals. Fisher's exact test is exact for the small
per-cell counts common in RAD panels, and the G-test is its
log-likelihood-ratio analogue. Figure \ref{fig:mode_candidates} compares
the candidate lists these choices produce on the real panels. For a marker
seen in \(x\) of \(m\) individuals in group 1 and \(y\) of \(f\) in group 2, we
keep the published chi-squared/Yates \cite{yates1934contingency}/Bonferroni route so that new results
remain comparable with prior RADSex studies. The p-value follows from the
identity
\begin{equation}
p = \mathrm{erfc}\Bigl(\sqrt{\frac{\chi^2}{2}}\Bigr)
\end{equation}
for one degree of freedom. Specializing the regularized lower gamma
\(P(1/2, \chi^2/2)\) and the relation \(\Gamma(1/2, z) = \sqrt{\pi} \cdot
\mathrm{erf}(\sqrt{z})\) (NIST DLMF 8.2.1 \cite{nist2024dlmf}; see also Abramowitz and Stegun \cite{abramowitz1964handbook}) yields the identity.\footnote{A mechanistic symbolic derivation of this identity, checked against reference chi-squared tail probabilities, is included with rsx (using SymPy \cite{meurer2017sympy}) and runs as part of the precision test suite.}

We also report the Bayes factor \cite{kass1995bayes} for the identical
2x2 table under a conjugate Beta-Binomial model. Let
\((a_1,b_1)\) and \((a_2,b_2)\) be the Beta shapes for the two group rates,
and \((a_0,b_0)\) those for the shared null rate. Then
\begin{equation}
  BF_{10} =
  \frac{B(x+a_1,m-x+b_1)B(y+a_2,f-y+b_2)B(a_0,b_0)}
       {B(a_1,b_1)B(a_2,b_2)B(x+y+a_0,m+f-x-y+b_0)}.
\end{equation}
This expression is the closed-form marginal-likelihood ratio; rsx requires
no simulation or numerical integration. The compatibility profile sets all
six shapes to one. Under that profile, Beta(1,1) assigns one
pseudocount to presence and one to absence, treats both outcomes
symmetrically, and leaves the prevalence uniform before the marker counts
are observed. We use it as a weak baseline for the Bayes factor rather than
as a taxon-specific prevalence estimate. Users may instead supply any
positive Beta shapes separately for the two alternatives and the shared null.

The reported sex-linkage posterior is a separate directional mixture, not
the Beta-Binomial Bayes factor converted to a probability. Its linked and
null prevalence families are independently configurable. A fixed family
evaluates the Bernoulli likelihood at a supplied probability; a Beta family
integrates that probability analytically with any positive shape parameters.
The two directional components describe an XY-like excess in group 1 and a
ZW-like excess in group 2. Their mixture weight is also configurable. Prior
odds \(\pi/(1-\pi)\) then give the posterior odds. The closed forms are given
in Appendix A.4.
The probability conversion is clamped when the absolute log-odds exceeds
20 to avoid floating-point overflow.

The compatibility defaults use fixed linked and null prevalences of 0.9 and
0.5, an equal directional mixture, and \(\pi=0.01\). They encode a screening design:
before examining a tag, one in one hundred tags is treated as linked, and a
linked marker may be absent from ten percent of individuals in the
heterogametic group. These values do not represent universal biological
frequencies. A study with known dropout, pedigree information, or validated
sex-linked loci should set the fixed probability or Beta shapes from those
controls and set \(\pi\) from the expected candidate burden. This explicit
prior surface follows the same user-controlled model-specification principle
as brms \cite{buerkner2017brms}, while retaining a small conjugate model rather
than introducing sampling or a general formula language. With no such calibration, the strict
Bonferroni result remains the primary call and the posterior serves only to
rank hypotheses. Our sensitivity grid spans
\(\pi\in\{0.001,0.005,0.01,0.02,0.05,0.1\}\) and
\(p_\text{sex}\in\{0.80,0.85,0.90,0.95\}\) under the fixed family, covering a hundred-fold change
in prior odds and linked-sex dropout from 5 to 20 percent.

A Bayes-factor threshold asks whether separate group prevalences describe
the 2x2 table better than one shared prevalence. The directional posterior
instead asks how strongly the counts resemble the specified XY or ZW
patterns under the chosen prior. On the \emph{N. rossii} panel, four hundred
rows exceed \(BF > 10\) but remain below posterior probability 0.9 and are
excluded from the sex-system call. The value 0.9 denotes posterior odds of
at least 9:1 and sets a candidate-ranking boundary; it was not optimized
against experimentally validated labels. Receiver-operating or
precision--recall calibration would require such labels, which these four
public panels do not provide for posterior-only markers. The output keeps
strict, posterior-supported, and Bayes-factor-only grades separate so a
study can choose its validation set and inspect the complete scores.

The model treats individuals as independent and requires correct
phenotypic-sex labels plus stable group-specific detection probabilities. Allele dropout,
restriction-site polymorphism, uneven coverage, related samples,
population structure, or batch effects can violate those assumptions and
inflate either direction of evidence. Depth filtering, sample-level PCA,
and sensitivity analysis can reveal some of these problems but cannot
remove model misspecification. Posterior-only rows therefore remain
computational hypotheses pending mapping, segregation, or experimental
validation.

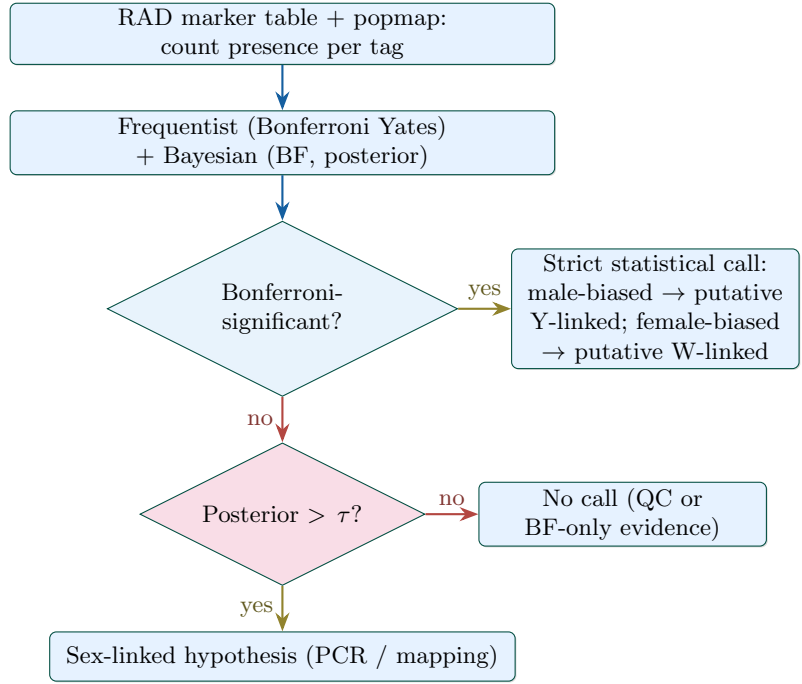
\begin{figure}[htbp]
\centering
\small
\tikzset{
  rsxbox/.style={draw=rsxteal, fill=rsxlight, rounded corners=2.5pt, align=center, text width=0.68\linewidth, minimum height=0.65cm, font=\small, inner sep=2pt, drop shadow={shadow xshift=0.5pt, shadow yshift=-0.5pt, opacity=0.3}},
  rsxdecision/.style={draw=rsxteal, fill=rsxsky!10, diamond, aspect=2, align=center, text width=0.22\linewidth, inner sep=2pt, font=\small},
  rsxarrow/.style={-{Stealth[length=2.5mm]}, thick, rsxsky!70!black},
  yesarrow/.style={-{Stealth[length=2.5mm]}, thick, rsxsunshine!60!black},
  noarrow/.style={-{Stealth[length=2.5mm]}, thick, rsxcoral!70!black}
}
\begin{tikzpicture}[node distance=0.6cm and 0.7cm]
\node[rsxbox, text width=0.54\linewidth] (input) {RAD marker table + popmap: count presence per tag};
\node[rsxbox, below=of input, text width=0.54\linewidth] (models) {Frequentist (Bonferroni Yates) + Bayesian (BF, posterior)};
\node[rsxdecision, below=of models] (strict) {Bonferroni-significant?};
\node[rsxbox, right=of strict, text width=0.28\linewidth] (confirm) {Strict statistical call: male-biased $\to$ putative Y-linked; female-biased $\to$ putative W-linked};
\node[rsxdecision, below=of strict, fill=rsxmagenta!15] (posterior) {Posterior $> \tau$?};
\node[rsxbox, right=of posterior, text width=0.28\linewidth] (negative) {No call (QC or BF-only evidence)};
\node[rsxbox, below=of posterior, text width=0.46\linewidth] (hypothesis) {Sex-linked hypothesis (PCR / mapping)};
\draw[rsxarrow] (input) -- (models);
\draw[rsxarrow] (models) -- (strict);
\draw[yesarrow] (strict) -- node[above, font=\scriptsize, text=rsxsunshine!40!black] {yes} (confirm);
\draw[noarrow] (strict) -- node[left, font=\scriptsize, text=rsxcoral!50!black] {no} (posterior);
\draw[noarrow] (posterior) -- node[above, font=\scriptsize, text=rsxcoral!50!black] {no} (negative);
\draw[yesarrow] (posterior) -- node[left, font=\scriptsize, text=rsxsunshine!40!black] {yes} (hypothesis);
\end{tikzpicture}
\caption{Biological interpretation of a RAD marker in rsx. Strict calls meet the matched RADSex statistical criterion; the posterior layer supplies lower-stringency hypotheses without altering the published panel-level classifications. Neither grade constitutes experimental validation.}
\label{fig:marker_decision_flow}
\end{figure}
\subsection{Biological inference outputs}
\label{sec:orgd76bad3}

Each rsx command and mode beyond the core RADSex set serves a specific
interpretation step in a RAD-seq sex-determination study. They map to the
following inference tasks:

\begin{table}[h]
\centering
\begin{tabular}{p{0.20\linewidth}p{0.25\linewidth}p{0.43\linewidth}}
\toprule
Inference task & rsx feature & Biological use \\
\midrule
Sequencing QC & \texttt{depth}, \texttt{freq}, PCA & choose depth thresholds; flag outlier libraries \\
Marker discovery & \texttt{distrib}, \texttt{signif} & test sex-biased marker presence/absence \\
Sex-linkage posterior & Bayes posterior output & classify markers as putative Y-linked, putative W-linked, or unsupported \\
Candidate validation & \texttt{subset} & inspect marker depths in individuals and outliers \\
Locus localization & \texttt{map} & test whether candidates cluster on a sex chromosome \\
Cross-run synthesis & \texttt{merge}, k-mer deduplication & combine lanes/panels and reduce redundant error variants \\
Reproducible analysis & Python bindings, C API & integrate outputs into notebooks and workflow languages \\
\bottomrule
\end{tabular}
\caption{Biological interpretation role of rsx features beyond RADSex CLI
compatibility.}
\label{tab:inference_roles}
\end{table}

The bounded-memory implementation matters for inference because rare
sex-linked RAD markers are sensitive to downsampling and marker-depth
thresholds. Keeping the full marker table preserves low-frequency
candidates for strict testing, posterior ranking, genome localization,
and sample-level quality control.

\begin{figure}[htbp]
\centering
\footnotesize
\tikzset{
  bioflow/.style={draw=rsxteal, fill=rsxlight, rounded corners=2.5pt, align=center, text width=0.26\linewidth, minimum height=0.65cm, inner sep=2pt},
  bioarrow/.style={-{Stealth[length=2.5mm]}, thick, rsxsky!70!black}
}
\begin{tikzpicture}[node distance=0.42cm]
\node[bioflow] (reads) {FASTQ reads + popmap};
\node[bioflow, right=of reads] (table) {Marker-by-individual depth table (mmap)};
\node[bioflow, right=of table] (counts) {Presence counts (males / females)};
\node[bioflow, below=of table] (qc) {Depth / freq / PCA QC};
\node[bioflow, below=of counts, fill=rsxsky!12] (models) {Bonferroni + BF + posterior};
\node[bioflow, below=of models, fill=rsxmagenta!10] (biology) {Putative Y-linked / W-linked / no call};
\node[bioflow, below=of biology] (follow) {Inspect, map, validate};
\draw[bioarrow] (reads) -- (table);
\draw[bioarrow] (table) -- (counts);
\draw[bioarrow] (table) -- (qc);
\draw[bioarrow] (counts) -- (models);
\draw[bioarrow] (qc) -- (models);
\draw[bioarrow] (models) -- (biology);
\draw[bioarrow] (biology) -- (follow);
\end{tikzpicture}
\caption{Biological workflow encoded by rsx. All paths remain bounded by $n_{\mathrm{ind}}$ or an explicit buffer; the output is an auditable sex-linked marker interpretation.}
\label{fig:biological_workflow}
\end{figure}
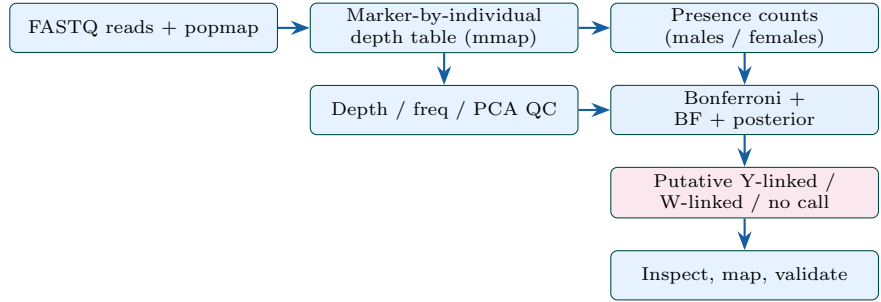
\subsection{2-bit DNA packing and k-mer deduplication}
\label{sec:orgb6003e1}

A RAD tag is a short DNA string, typically 80 to 100 bases. DNA has only
four letters, so each base needs just two bits (A=00, C=01, G=10, T=11),
and rsx packs each tag into this form before counting it. A 100-base tag
then occupies 26 bytes instead of 100, a four-fold saving in the lookup
table that contains one entry per distinct tag.\footnote{In rsx this table is a
hash map keyed by the packed tag: an \texttt{AHashMap} on the serial path or a
sharded \texttt{DashMap} on the parallel ingestion path.} When a panel contains
millions of distinct tags, that table dominates memory, so the saving can
decide whether the run fits in RAM.

A second, optional step removes the near-identical tags that sequencing
error creates. Two reads that differ by a single miscalled base come from
the same locus, yet they pack to different keys and count as two tags. To
catch them, rsx can group tags by a min-hash. It slides a window of length
\(k\) along each tag; reduces every window to whichever of the window and its
reverse complement is lexicographically smaller (the canonical k-mer), so
that a tag and its reverse-complement read agree; and signs the whole tag
with the smallest hash value among those windows. Tags that share a
signature may come from related tags, so merging them can collapse error
variants. Two tags merge when their minimum canonical k-mer is equal, which is
not the same as being one base apart: a substitution falling inside the minimum
window changes the signature and the pair does not merge, while two unrelated
tags that happen to share a minimum window do. The option is off by default and
unsuitable when marker identity has to be exact. We measure how often it
recovers a one-base variant on the reference tables below. Moeckel et al. \cite{moeckel2024survey} survey the
wider family of low-memory, streaming k-mer methods that make such
packing and sketching practical at sequencing scale.
\subsection{Streaming design and statistical triage}
\label{sec:orgf6bf35f}

A marker table is a matrix: one row per RAD tag, one column per
individual, with each cell holding the read depth of that tag in that
sample. On the literature panels studied here the matrix runs to tens of
millions of rows and only tens to a few hundred columns, so a single table
contains billions of cells and a plain in-memory copy reaches tens of
gigabytes. RADSex constructs and stores the whole matrix, which is why
its table-building commands run out of memory on large panels. rsx instead
implements each marker-table operation as a streaming kernel with an
explicit storage bound, so memory scales with the number of individuals,
or with a buffer the user sets, and not with the number of markers. The
design rests on four recurring elements:
\begin{itemize}
\item streaming producer-consumer pipelines that never hold the full marker
table in RAM,
\item two-pass memory-mapped scans that apply global multiple-testing
corrections without materialising the whole table,
\item external sorting for order statistics and table merging: when the
in-memory buffer fills, sorted runs spill to compressed temporary files
and a final k-way merge produces the output, so peak memory stays at the
user-chosen buffer no matter how large the input,
\item hybrid frequentist/Bayesian triage that combines strict error
control with a posterior probability of sex-linkage.
\end{itemize}

Figure \ref{fig:algo_flow} summarizes the data flow, and Algorithms
\ref{alg:triage} and \ref{alg:twopass} state the corresponding decision
rules. Except where we state a numerical approximation explicitly, the
streaming kernels compute the same statistic as the corresponding
full-table algorithm. The stated statistical model and
thresholds govern marker calls rather than truncation of the marker table.

\begin{figure}[htbp]
\centering
\footnotesize
\tikzset{
  rsxbox/.style={draw=rsxteal, fill=rsxlight, rounded corners=2.5pt, align=center, text width=0.30\linewidth, minimum height=0.8cm, font=\small, inner sep=2pt, drop shadow={shadow xshift=0.5pt, shadow yshift=-0.5pt, opacity=0.3}},
  rsxsmall/.style={draw=rsxteal, fill=rsxsky!8, rounded corners=2pt, align=center, text width=0.155\linewidth, minimum height=0.9cm, font=\scriptsize, inner sep=1pt},
  rsxarrow/.style={-{Stealth[length=1.5mm]}, thick, rsxteal}
}
\begin{tikzpicture}[node distance=0.5cm and 0.12cm]
\node[rsxbox] (fastq) {FASTQ (needletail) + popmap};
\node[rsxbox, below=of fastq] (proc) {process (rayon + ahash + 2-bit pack) \\ $O(n_{\mathrm{ind}})$};
\node[rsxbox, below=of proc] (markers) {markers.tsv (mmap, 2-bit keys)};
\node[rsxsmall, below=1.0cm of markers, fill=rsxmagenta!20] (signif) {signif/subset \\ bitset two-pass \\ $O(n_{\mathrm{ind}})$};
\node[rsxsmall, left=of signif] (distrib) {distrib \\ $O(n_{\mathrm{ind}})$};
\node[rsxsmall, left=of distrib, fill=rsxcoral!15] (triage) {triage \\ BF + posterior \\ $O(n_{\mathrm{ind}})$};
\node[rsxsmall, right=of signif, fill=rsxcoral!15] (depth) {depth/merge \\ ext. sort \\ $O(\mathrm{buffer})$};
\node[rsxsmall, right=of depth] (map) {map \\ minimap2 \\ $O(\mathrm{genome})$};
\draw[rsxarrow] (fastq) -- (proc);
\draw[rsxarrow] (proc) -- (markers);
\draw[rsxarrow] (markers.south) -- (triage.north);
\draw[rsxarrow] (markers.south) -- (distrib.north);
\draw[rsxarrow] (markers.south) -- (signif.north);
\draw[rsxarrow] (markers.south) -- (depth.north);
\draw[rsxarrow] (markers.south) -- (map.north);
\node[font=\scriptsize, below=0.45cm of signif, text width=0.92\linewidth, align=center] {Every path bounded by $n_{\mathrm{ind}}$ or explicit buffer; no full $n_{\mathrm{markers}} \times n_{\mathrm{ind}}$ materialisation except optional BH/FDR ranking.};
\end{tikzpicture}
\caption{Bounded-memory data flow in rsx. Parallel ingestion produces a 2-bit-packed memory-mapped table; downstream commands are bounded by $n_{\mathrm{ind}}$ or by an explicit user buffer.}
\label{fig:algo_flow}
\end{figure}

Algorithm \ref{alg:triage} defines the transformation from per-marker
presence counts to the three evidence labels used throughout the results:
strict statistical calls, posterior-supported hypotheses, and
Bayes-factor-only rows that do not pass the posterior threshold. The
algorithm first obtains the Bonferroni denominator by a fast count of the
table. It then streams the table a second time, computes the
bitset-masked group counts, applies the frequentist test, evaluates the
Bayes factor and posterior mixture, and emits markers that cross at least one
threshold together with the label specifying which threshold was crossed. If
none crosses a threshold, the highest-posterior marker is retained as an
\texttt{exploratory} row rather than a call; ties are resolved by Bayes factor and
then p-value. The same threshold rule produces the reported numbers and
biological classifications for the four literature panels.

\begin{algorithm}[htbp]
\caption{Hybrid strict + Bayesian marker triage (core of rsx-triage)}
\label{alg:triage}
\begin{algorithmic}[1]
\Require Marker table $T$, popmap $P$, $\tau_p$, $\tau_{post}$, $\tau_{BF}$, $\pi$, $p_{sex}$
\State Compute group totals and GroupMask bitsets from $P$
\State $N \gets$ fast count of markers in $T$ (Pass 1, sequence-free)
\State $\tau_{bonf} \gets \tau_p / N$
\State $best \gets \varnothing$; $n_{called} \gets 0$
\For{each marker $m$ in $T$ (Pass 2, streaming)}
  \State $g_1, g_2 \gets$ presence counts via masked bitsets
  \State $p \gets$ chi\_squared\_yates\_p($g_1,g_2,\dots$)
  \State $strict \gets (p < \tau_{bonf})$
  \State $BF \gets$ bayes\_factor\_2x2($g_1,g_2,\dots$)   \Comment{Beta-Binomial marginal ratio}
  \State $post \gets$ posterior\_sex\_linked($g_1,g_2,\dots,\pi,p_{sex}$) \Comment{two-dir mixture + null}
  \If{$strict \lor post > \tau_{post} \lor BF > \tau_{BF}$}
    \State Emit $m$ with (strict, post, BF, class label)
    \State $n_{called} \gets n_{called} + 1$
  \ElsIf{$best = \varnothing$ or $(post,BF,-p)_m > (post,BF,-p)_{best}$}
    \State $best \gets m$
  \EndIf
\EndFor
\If{$n_{called}=0$ and $best \ne \varnothing$}
  \State Emit $best$ with class = exploratory
\EndIf
\end{algorithmic}
\end{algorithm}

Algorithm \ref{alg:twopass} gives the Bonferroni scan used by four commands:
\texttt{signif}, \texttt{subset}, \texttt{map} and \texttt{triage}.

\begin{algorithm}[htbp]
\caption{Two-pass streaming Bonferroni (bounded memory, no full table in RAM)}
\label{alg:twopass}
\begin{algorithmic}[1]
\Require Table $T$ (mmap), threshold $\tau$, popmap masks
\State $N \gets$ fast line count of $T$ (memory-map, no sequence parse) \Comment{Pass 1}
\State $\tau' \gets \tau / N$
\For {marker $m$ in $T$ (Pass 2, re-mmap, kernel cache hits)}
  \State $g_1, g_2 \gets$ bitset-masked presence counts (O(1) per marker)
  \State $p \gets$ chi\_squared\_yates\_p($g_1, g_2, \dots$)
  \If {$p < \tau'$}
    \State Emit $m$ (with corrected $p$)
  \EndIf
\EndFor
\end{algorithmic}
\end{algorithm}

The same principle applies to the external-sort median and the streaming
Gram-PCA accumulation. For PCA, the right singular vectors of the
centered marker-by-sample matrix are exactly the eigenvectors of the
sample Gram matrix (mode-2 unfolding). Appendix A gives the derivation and
the resulting \(O(n_{\mathrm{ind}}^2)\) memory bound.\footnote{rsx includes this
derivation as an executable symbolic script, so the bound can be re-checked
mechanically rather than taken on trust.}
These statements provide an implementation-independent specification of
the performance and numerical claims.
\subsection{Bitset-masked two-pass Bonferroni scan}
\label{sec:org60d1d0d}

Counting how many males and how many females carry a marker is the inner
loop of every significance test, run once per marker over millions of
markers, so rsx makes it cheap with bit operations. It records a marker's
presence across the \(n\) individuals as a bitset: a string of \(n\) bits
(packed into one or a few 64-bit machine words) in which bit \(i\) is 1 when
individual \(i\) carries the tag. From the population map it constructs two fixed
group masks just once, \(\mathrm{mask}_1\) with a 1 in every male position
and \(\mathrm{mask}_2\) in every female position. For a marker with presence
bits \(m\), the number of males carrying it is then
\(g_1 = \mathrm{popcount}(\mathrm{mask}_1 \land m)\): the bitwise AND
(\(\land\)) keeps only the bits set in both the mask and the marker, that is,
the males that carry the tag, and \(\mathrm{popcount}\) (population count), a
single CPU instruction that returns the number of 1-bits in a machine word,
counts how many of those bits remain set. The female count \(g_2\)
follows the same way. Each marker therefore costs several word-sized
AND and popcount operations rather than a loop over all \(n\) samples, and
the only state in memory is the \(n\) presence bits plus a few accumulators.

\begin{figure}[htbp]
\centering
\scriptsize
\tikzset{
  bitbox/.style={draw=rsxteal, fill=rsxlight, rounded corners=2.5pt, align=center, text width=0.36\linewidth, minimum height=0.6cm, inner sep=2pt},
  bitarrow/.style={-{Stealth[length=2mm]}, thick, rsxsky!70!black}
}
\begin{tikzpicture}[node distance=0.26cm]
\node[bitbox] (mmap) {mmap markers.tsv (Pass 1: count $N$)};
\node[bitbox, below=of mmap] (mask) {Build 64-bit GroupMask bitsets (once)};
\node[bitbox, below=of mask] (pass2) {Pass 2 (re-mmap, page cache): $g_1 = \mathrm{popcount}(\mathrm{mask}_1 \land m)$, $g_2 = \mathrm{popcount}(\mathrm{mask}_2 \land m)$};
\node[bitbox, below=of pass2] (pval) {$p = \chi^2_{\mathrm{Yates}}(g_1,g_2)$; emit if $p < \tau/N$};
\draw[bitarrow] (mmap) -- (mask);
\draw[bitarrow] (mask) -- (pass2);
\draw[bitarrow] (pass2) -- (pval);
\node[font=\tiny, below=1.4cm of pval, text width=0.82\linewidth, align=center, fill=rsxsunshine!20, rounded corners=2pt, draw=rsxteal, inner sep=2pt] {Only $n_{\mathrm{ind}}$ bits + accumulators in RAM. Pass 2 reuses cached pages when available. Used by signif and triage as well as subset/map.};
\end{tikzpicture}
\caption{Bitset-masked two-pass Bonferroni scan. The global correction is applied while retaining only $n_{\mathrm{ind}}$ bits and accumulators in RAM.}
\label{fig:bitset_twopass}
\end{figure}
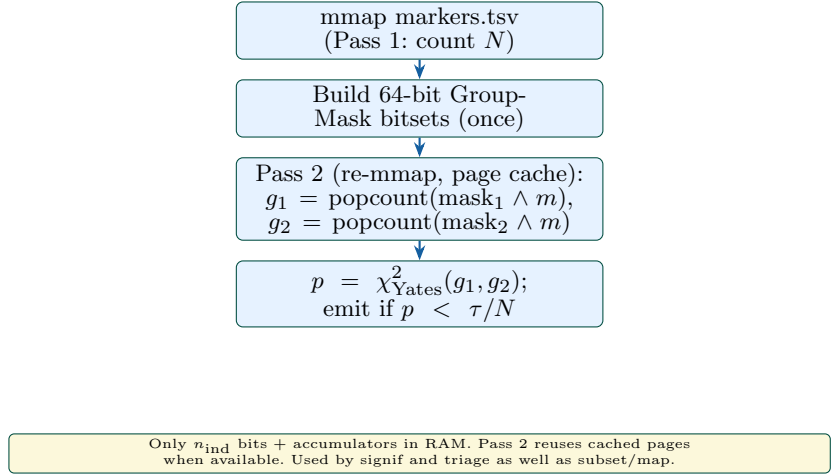

Algorithm \ref{alg:posterior} gives the posterior update used by the
Bayesian layer. We spell out the modeling assumptions because downstream
biological decisions use the posterior probability rather than a software
score.

Each RAD tag's presence or absence in the two phenotypic-sex groups gives a
series of labelled Bernoulli observations. A fixed prevalence family evaluates
their likelihood at a user-supplied probability. A Beta family instead
integrates that probability and yields a closed ratio of Beta functions. The
implementation evaluates the ratio through log-gamma, so neither family needs
sampling or numerical quadrature.

The unknown sex-determination system may follow male heterogamety (XY) or
female heterogamety (ZW). For one direction the concordance count is the number
present in group 1 plus the number absent in group 2; the other reverses those
roles. A configurable weight averages the two directional likelihoods. The
null uses the pooled presence count under its own independently configured
prevalence family. Prior log-odds \(\ln(\pi/(1-\pi))\) then convert that evidence
ratio to a posterior probability. The compatibility profile fixes the linked
and null prevalences at 0.9 and 0.5 and assigns the directions equal weight;
Beta-integrated profiles use the same calculation with supplied positive shape
parameters. The guard at \(|\log\text{-odds}|>20\) prevents numerical overflow.

\begin{algorithm}[htbp]
\caption{Configurable directional posterior. $E(k,N\mid\mathcal P)$ is either the fixed-prevalence Bernoulli likelihood or the Beta-integrated likelihood defined in Appendix A.4.}
\label{alg:posterior}
\begin{algorithmic}[1]
\State $N \gets n_1+n_2$
\State $k_1 \gets g_1+(n_2-g_2)$; $k_2 \gets (n_1-g_1)+g_2$; $k_0 \gets g_1+g_2$
\State $ll_1 \gets \log E(k_1,N\mid\mathcal P_L)$; $ll_2 \gets \log E(k_2,N\mid\mathcal P_L)$
\State $ll_{linked} \gets \mathrm{logsumexp}(ll_1+\ln w,ll_2+\ln(1-w))$
\State $ll_{null} \gets \log E(k_0,N\mid\mathcal P_0)$
\State $\mathrm{logodds} \gets ll_{linked}-ll_{null}+\ln(\pi/(1-\pi))$
\State $\mathrm{post} \gets 1 / (1 + \exp(-\mathrm{logodds}))$ (with guard $| \mathrm{logodds} | > 20$)
\end{algorithmic}
\end{algorithm}

\begin{figure}[htbp]
\centering
\includegraphics[width=0.92\linewidth]{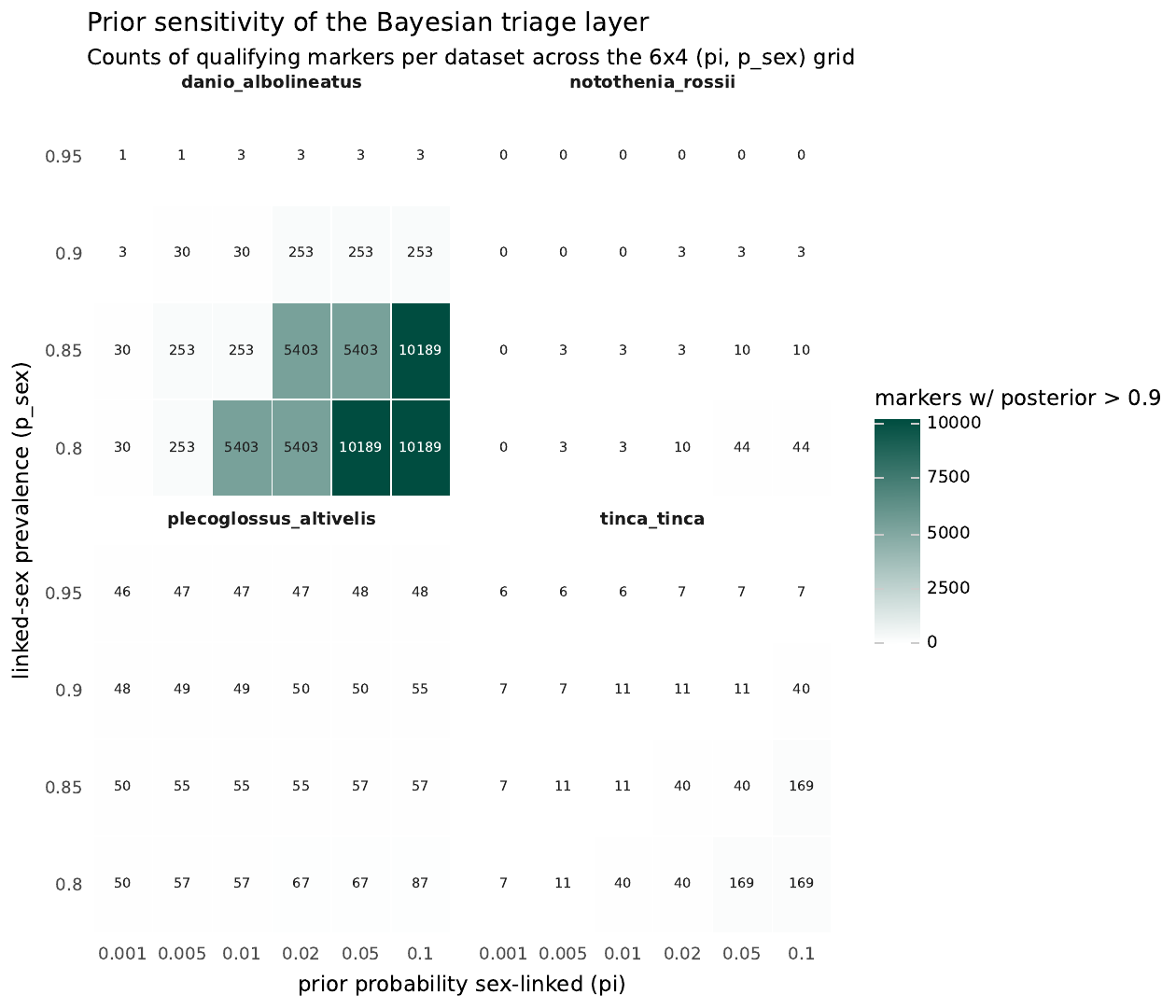}
\caption{Prior sensitivity of the Bayesian triage layer on the four real
literature marker tables. The Bayesian call depends on two inputs the user
sets: $\pi$, the prior probability specified for any given RAD tag to be
sex-linked (horizontal coordinate), and $p_{\mathrm{sex}}$, the expected detection
probability of a linked marker in the heterogametic sex (vertical coordinate). Each
cell reports the number of markers with posterior probability above 0.9
when \texttt{rsx triage} is re-run at that parameter pair. The wide count
range for /D. albolineatus/ shows why posterior-only rows require explicit
prior reporting and external validation.}
\label{fig:prior_sensitivity}
\end{figure}

Algorithm \ref{alg:posterior} returns the model's answer to the question:
given a prior belief that a random RAD tag has probability
\(\pi\) of being sex-linked, and given the male and female counts observed
for this tag, what is the updated probability that the tag is sex-linked?
That probability decides whether
the evidence for a particular tag justifies the next
round of PCR or genome mapping. Repeating this computation across
an entire table under different values of \(\pi\) produces the
prior-sensitivity analysis in Figure \ref{fig:prior_sensitivity}.

Together, the data-flow summary, formal algorithms, and SymPy
derivation scripts give an implementation-independent description of the
performance and statistical claims in this paper.
\section{Results}
\label{sec:org77753e0}

\subsection{Performance benchmarks}
\label{sec:orgf99e549}

We compared rsx against RADSex v1.2.0 on synthetic RAD-seq-like
datasets generated by the reproducibility package script
\texttt{scripts/generate\_data.py} (vendored in the archive) and on downloaded
RAD-seq datasets from the RADSex literature workflow. The synthetic suite
is a compact regression benchmark; the literature suite contains the
biological performance numbers. The C++ implementation requires 1.558 s
in total across the 19 paired synthetic timings; rsx requires 0.780 s
(2.0-fold speedup).

\begin{table}[h]
\centering
\begin{tabular}{lrrr}
\toprule
Command/scale & RADSex (s) & rsx (s) & Speedup \\
\midrule
small freq & 0.016 & 0.004 & 4.0x \\
small distrib & 0.015 & 0.003 & 5.0x \\
small signif & 0.015 & 0.003 & 5.0x \\
medium process & 0.058 & 0.038 & 1.5x \\
medium map & 0.407 & 0.228 & 1.8x \\
large freq & 0.114 & 0.044 & 2.6x \\
large depth & 0.240 & 0.067 & 3.6x \\
large distrib & 0.138 & 0.057 & 2.4x \\
large signif & 0.189 & 0.137 & 1.4x \\
large subset & 0.170 & 0.118 & 1.4x \\
\bottomrule
\end{tabular}
\caption{Representative benchmark timings. The complete repository CSV
contains 19 paired command/scale measurements.}
\end{table}

After one in-process initialization, the CUDA p-value backend crossed 10-fold total speedup at 100,000 markers on an NVIDIA A100 and at 1,000,000 markers on an RTX 5070. At one million markers the respective speedups were 29.86-fold and 14.43-fold.

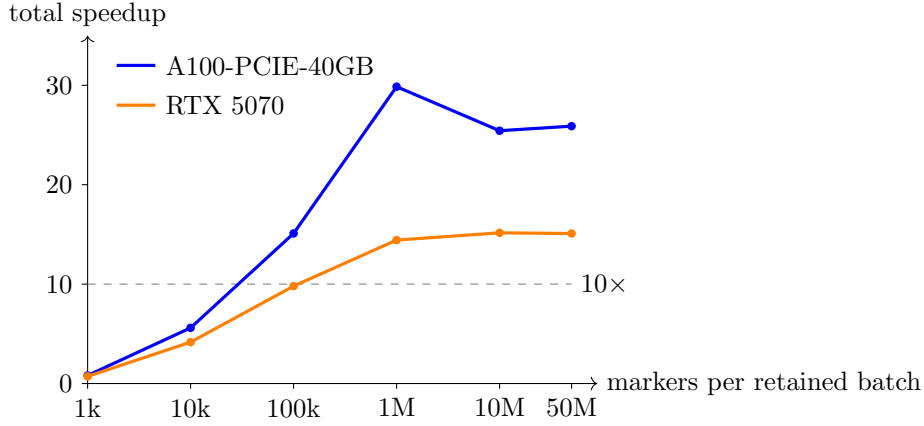
\begin{figure}[htbp]
\centering
\begin{tikzpicture}[x=1.3620cm,y=0.1314cm]
  \draw[->] (0,0) -- (4.9490,0) node[right] {markers per retained batch};
  \draw[->] (0,0) -- (0,35) node[above] {total speedup};
  \foreach \x/\label in {0.0000/{1k}, 1.0000/{10k}, 2.0000/{100k}, 3.0000/{1M}, 4.0000/{10M}, 4.6990/{50M}} {
    \draw (\x,0) -- (\x,-0.5250) node[below] {\label};
  }
  \foreach \y in {0, 10, 20, 30} {
    \draw (0,\y) -- (-0.0564,\y) node[left] {\y};
  }
  \draw[dashed,gray] (0,10) -- (4.6990,10) node[right,black] {$10\times$};
  \draw[very thick,blue] plot coordinates {(0.0000,0.810) (1.0000,5.600) (2.0000,15.100) (3.0000,29.860) (4.0000,25.430) (4.6990,25.890)};
  \fill[blue] (0.0000,0.810) circle (1.6pt);
  \fill[blue] (1.0000,5.600) circle (1.6pt);
  \fill[blue] (2.0000,15.100) circle (1.6pt);
  \fill[blue] (3.0000,29.860) circle (1.6pt);
  \fill[blue] (4.0000,25.430) circle (1.6pt);
  \fill[blue] (4.6990,25.890) circle (1.6pt);
  \draw[very thick,orange] plot coordinates {(0.0000,0.720) (1.0000,4.160) (2.0000,9.800) (3.0000,14.430) (4.0000,15.160) (4.6990,15.090)};
  \fill[orange] (0.0000,0.720) circle (1.6pt);
  \fill[orange] (1.0000,4.160) circle (1.6pt);
  \fill[orange] (2.0000,9.800) circle (1.6pt);
  \fill[orange] (3.0000,14.430) circle (1.6pt);
  \fill[orange] (4.0000,15.160) circle (1.6pt);
  \fill[orange] (4.6990,15.090) circle (1.6pt);
  \draw[very thick,blue] (0.2819,32) -- (0.6579,32)
    node[right,black] {A100-PCIE-40GB};
  \draw[very thick,orange] (0.2819,28) -- (0.6579,28)
    node[right,black] {RTX 5070};
\end{tikzpicture}
\caption{Five-repetition CUDA p-value crossover on both devices, including
host--device transfers and synchronization. The dashed line marks ten-fold
total speedup. The accelerated region is the batched Yates statistic and erfc
evaluation; marker parsing, group counting, correction, filtering, and file
output remain on the CPU and are not included in these batch speedups. Drawn
from the same summary as Tables \ref{tab:cuda_elja} and
\ref{tab:cuda_devices}.}
\label{fig:cuda_crossover}
\end{figure}

The CUDA crossover benchmark evaluates the chi-squared p-value batch in one
process, in increasing sizes from 1,000 to 50 million markers, and repeats the
matrix five times. The first invocation initializes CUDA and compiles the
kernel. Four subsequent repetitions reuse the loaded kernel and the largest
page-locked result allocation. Table \ref{tab:cuda_elja} reports medians across
all five repetitions from 100,000 markers upward, the range that contains the
crossover; Figure \ref{fig:cuda_crossover} draws every size. The one-time setup occurs in only the first sample and
therefore does not determine the median. Times include both transfer directions
and the kernel.

\begin{table}[h]
\centering
\begin{tabular}{rrrrr}
\toprule
Markers & CPU (ms) & CUDA total (ms) & Kernel speedup & Total speedup \\
\midrule
100,000 & 4.539 & 0.301 & 205.30x & 15.10x \\
1,000,000 & 34.680 & 1.161 & 562.74x & 29.86x \\
10,000,000 & 357.172 & 14.046 & 717.46x & 25.43x \\
50,000,000 & 1780.238 & 68.758 & 739.51x & 25.89x \\
\bottomrule
\end{tabular}
\caption{Five-repetition CUDA crossover benchmark on an NVIDIA
A100-PCIE-40GB. Total time covers both transfer directions and synchronization
as well as kernel execution and retained-buffer bookkeeping. Medians run over
all five repetitions; the one-time setup falls in a single sample and so cannot
move a median of five. The largest absolute CPU/GPU p-value difference was
$1.1\times10^{-16}$; the 100,000-marker batch is the first median total speedup
above ten-fold.}
\label{tab:cuda_elja}
\end{table}

The same matrix was run on Terra with an NVIDIA GeForce RTX 5070. Table
\ref{tab:cuda_devices} compares the critical size and retained-batch behavior
on both machines. The A100 exceeded ten-fold total speedup at 100,000 markers
and the RTX 5070 at one million; both remained above ten-fold through the
largest tested batch.

\begin{table*}[t]
\centering
\small
\setlength{\tabcolsep}{4pt}
\begin{tabular}{lrrrrrr}
\toprule
Device & First $>10$x & 1M & 10M & 50M & H2D & D2H \\
\midrule
A100-PCIE-40GB & 100,000 & 29.86x & 25.43x & 25.89x & 9.67 & 16.03 \\
RTX 5070 & 1,000,000 & 14.43x & 15.16x & 15.09x & 19.56 & 28.58 \\
\bottomrule
\end{tabular}
\caption{CUDA critical size and retained-batch medians on the NVIDIA
A100-PCIE-40GB (Elja) and GeForce RTX 5070 (Terra). The three marker-count
columns give total speedup; H2D and D2H are GB/s at 50 million markers. The
data centre card crosses ten-fold an order of magnitude earlier because the
consumer card lacks full-rate binary64 arithmetic.}
\label{tab:cuda_devices}
\end{table*}

The same GPU executable was also run on the
334,411-locus \emph{Rana berlandieri} marker table. Both backends tested 311,313
markers and produced files with the same SHA256 digest on both machines. On
Elja, that one-shot command took 0.29~s on the CPU and 1.49~s with a cold CUDA
process; the corresponding Terra times were 0.157 and 0.723~s. These runs
establish output equivalence but not acceleration at that scale. The crossover
claim applies to retained, repeated batches rather than short independent
commands.

We also ran a five-repetition resource matrix on the same two machines, this
time on their CPUs. The Terra run is labelled \texttt{terra-interactive} in the
results. Both hosts used the pinned rsx and
RADSex v1.2.0 binaries and the same generated inputs: marker-table sizes
1,000/10,000/100,000 paired with 10/20/40 individuals.
Table commands use each program's normal thread configuration because
they do not expose the \texttt{process -T} control. The evidence files record
that setting as \texttt{default} rather than assigning an unsupported thread
count.

\begin{table*}[t]
\centering
\footnotesize
\setlength{\tabcolsep}{3pt}
\begin{tabular}{lllrrrr}
\toprule
Host & Implementation & Command & Wall (ms) & User (ms) & System (ms) & Max RSS (MiB) \\
\midrule
Elja & RADSex & depth   & $226.2 \pm 6.4$ & 252.7 & 14.4 & 18.6 \\
Elja & RADSex & distrib & $172.4 \pm 7.6$ & 211.9 & 7.0  & 15.1 \\
Elja & RADSex & freq    & $108.1 \pm 4.1$ & 108.5 & 7.0  & 15.1 \\
Elja & RADSex & signif  & $234.2 \pm 4.7$ & 268.6 & 18.7 & 22.0 \\
Elja & rsx    & depth   & $32.5 \pm 4.0$  & 164.8 & 41.2 & 40.8 \\
Elja & rsx    & distrib & $13.8 \pm 1.5$  & 97.9  & 5.2  & 15.1 \\
Elja & rsx    & freq    & $14.4 \pm 1.6$  & 111.6 & 4.9  & 15.1 \\
Elja & rsx    & signif  & $129.6 \pm 3.4$ & 114.2 & 7.0  & 22.9 \\
\midrule
Terra & RADSex & depth   & $316.4 \pm 82.8$ & 342.1 & 45.5 & 17.0 \\
Terra & RADSex & distrib & $241.9 \pm 25.5$ & 253.3 & 30.8 & 14.6 \\
Terra & RADSex & freq    & $127.7 \pm 7.5$  & 108.1 & 36.6 & 14.6 \\
Terra & RADSex & signif  & $357.4 \pm 47.9$ & 386.2 & 75.2 & 23.6 \\
Terra & rsx    & depth   & $71.2 \pm 12.6$  & 141.1 & 82.0 & 39.3 \\
Terra & rsx    & distrib & $20.5 \pm 3.0$   & 58.6  & 18.8 & 24.4 \\
Terra & rsx    & freq    & $20.3 \pm 3.1$   & 55.9  & 19.7 & 24.2 \\
Terra & rsx    & signif  & $208.1 \pm 44.5$ & 168.9 & 26.5 & 25.0 \\
\bottomrule
\end{tabular}
\caption{Five-run measurements on the 100,000-marker synthetic table.
Wall time is the median with sample standard deviation; user time,
system time, and maximum resident set size are medians. The complete raw
and summary CSV files retain each scale with its repetition and exit status.}
\label{tab:response_resources}
\end{table*}

The same measurement protocol was applied to the four deposited literature
tables, spanning 6.93--29.36 million markers. Table
\ref{tab:literature_resources} reports five-run medians for the two commands
shared by every panel. rsx was faster in all eight comparisons, with speedups
from 1.23-fold to 18.95-fold. Its maximum RSS was 1.22--6.01~GiB because Linux
counts the scanned, file-backed mapping while those pages remain resident;
RADSex reported 16.7--17.1~MiB. These measurements support the timing claim
and the design that bounds writable allocations, but they do not support a claim
that rsx always has a smaller process RSS. Raw per-run values, exit codes and
hardware metadata are tracked in \texttt{results/response/terra-literature} at
reproducibility commit \texttt{77e70c5}.

\begin{table*}[t]
\centering
\scriptsize
\setlength{\tabcolsep}{2pt}
\begin{tabular}{lrlrrrrr}
\toprule
Dataset & Markers & Command & RADSex & rsx & Speedup & RADSex RSS & rsx RSS \\
\midrule
\emph{D. albolineatus} & 29.36 & depth  & 45.71 & 37.09 & 1.23x  & 16.7 & 6151.3 \\
                   &       & signif & 71.22 & 30.57 & 2.33x  & 16.7 & 6154.2 \\
\emph{N. rossii}        & 11.34 & depth  & 14.90 & 5.80  & 2.57x  & 16.7 & 2036.2 \\
                   &       & signif & 21.67 & 8.82  & 2.46x  & 16.7 & 2028.8 \\
\emph{P. altivelis}     & 7.34  & depth  & 12.53 & 0.73  & 17.23x & 17.1 & 1658.6 \\
                   &       & signif & 16.02 & 11.67 & 1.37x  & 16.7 & 1635.8 \\
\emph{T. tinca}         & 6.93  & depth  & 9.46  & 0.50  & 18.95x & 16.7 & 1263.9 \\
                   &       & signif & 13.72 & 4.93  & 2.79x  & 16.7 & 1254.2 \\
\bottomrule
\end{tabular}
\caption{Five-run medians on deposited literature marker tables at minimum
depth 10. Marker counts are millions, wall times are seconds, and RSS values
are MiB. Maximum RSS is the child-process value returned by \texttt{wait4} and
includes resident file-backed mappings. The result archive retains wall,
user and system times with sample standard deviations for every row.}
\label{tab:literature_resources}
\end{table*}

The process scaling experiment uses the medium FASTQ input (20
individuals, 2,000 reads per individual) and sets \texttt{-T} to 1, 2, 4, 8,
and 16 for both programs. Table \ref{tab:thread_scaling} reports the
endpoints. Neither implementation scales linearly on this compact input:
at 16 threads, rsx reaches 4.79-fold speedup on Elja and 4.43-fold on
Terra, corresponding to 29.9 and 27.7 percent parallel efficiency.

\begin{table}[h]
\centering
\begin{tabular}{llrrrr}
\toprule
Host & Implementation & 1 thread (ms) & 16 threads (ms) & Speedup & Efficiency \\
\midrule
Elja  & RADSex & 140.17 & 32.23 & 4.35 & 27.2\% \\
Elja  & rsx    & 105.86 & 22.10 & 4.79 & 29.9\% \\
Terra & RADSex & 227.28 & 51.95 & 4.37 & 27.3\% \\
Terra & rsx    & 149.27 & 33.71 & 4.43 & 27.7\% \\
\bottomrule
\end{tabular}
\caption{Median wall time and parallel efficiency for five repeated
=process= runs. Efficiency is speedup divided by 16. Intermediate
2-, 4-, and 8-thread measurements are retained in the result CSV files.}
\label{tab:thread_scaling}
\end{table}

These generated inputs make the matrix quick enough to repeat and expose
host and scheduler effects. They are regression measurements, not a
replacement for the four literature-panel runs below.

The k-mer deduplication audit uses a seeded reservoir sample of 250 tags
from each deposited literature marker table. For each sampled tag and
each value \(k\in\{8,16,24,32\}\), the audit generates all three
substitutions at every base and checks whether the mutated tag retains the
reference tag's minimum canonical k-mer signature. Each cell in Table
\ref{tab:kmer_recall} therefore contains 70,500 mutation cases; the full
table contains 1,128,000 cases.

\begin{table}[h]
\centering
\begin{tabular}{lrrrr}
\toprule
Panel & $k=8$ & $k=16$ & $k=24$ & $k=32$ \\
\midrule
\emph{D. albolineatus} & 88.91\% & 80.69\% & 72.47\% & 64.19\% \\
\emph{N. rossii}       & 88.64\% & 80.48\% & 72.35\% & 64.21\% \\
\emph{P. altivelis}    & 87.70\% & 78.93\% & 71.16\% & 63.16\% \\
\emph{T. tinca}        & 88.33\% & 80.12\% & 71.91\% & 63.68\% \\
\bottomrule
\end{tabular}
\caption{Single-substitution recall of the optional minimum-canonical-k-mer
grouping rule. The option remains disabled by default.}
\label{tab:kmer_recall}
\end{table}

Recall decreases as \(k\) increases because a substitution affects a larger
fraction of the available windows. Even at \(k=8\), 11.1--12.3 percent of
the one-base variants receive a different signature. This measured result
rules out an edit-distance guarantee. The audit does not estimate false
merges between unrelated tags, and its seeded sample does not establish a
taxon-independent recall rate.

Commands that parse tables and count groups produce the largest gains.
The smaller gains for signif and subset at large scale follow from the
two-pass design, which keeps memory bounded for Bonferroni and uncorrected
output. The \texttt{map} command localizes candidate markers on a reference genome. The
original RADSex pipeline aligns them with BWA-MEM \cite{li2009bwa}, whereas rsx calls
minimap2 \cite{li2018minimap2}, a faster modern aligner. The two run
different alignment algorithms, so the \texttt{map} timing is not the cost of an
identical computation; it measures practical throughput for the
localization step. Li et al. \cite{li2023pcaone} and
Urgese et al. \cite{urgese2020bioseqzip} discuss out-of-core PCA and
external-sort precedents for large genomics matrices.
\subsection{Literature-derived biological benchmarks}
\label{sec:orgded2b1e}

The biological benchmark corpus comes from published RADSex resources.
The reproducibility archive contains the manifest of the 15 public RADSex
workflow datasets \cite{feron2021radsex} (BioProject PRJNA548074) plus
the \emph{Oryzias latipes} medaka example from Wilson et al.
\cite{wilson2014wild} (BioProject PRJNA253959). The RADSex workflow
executes the process/distrib/signif/freq/depth command family at minimum depths
\(\{1,2,5,10\}\); the medaka case also runs map and subset against the
published reference.

We downloaded and executed four RADSex workflow datasets from FASTQ processing
through marker-table analysis:
\emph{Danio albolineatus} (58 samples, 29 357 812 markers), \emph{Notothenia rossii}
(42 samples, 11 337 736 markers), \emph{Plecoglossus altivelis} (65 samples, 7
338 124 markers), and \emph{Tinca tinca} (43 samples, 6 932 387 markers). The
runs cover 41.9 billion sequenced bases in total. All performance numbers
below report compute time only (FASTQ processing plus marker-table
analysis). The reproducibility table retains download time, but we make
no performance claim from it.

We chose these inputs for their published biological outcomes. In
the RADSex study, \emph{P. altivelis} recovered a known XX/XY system with 47
sex-associated markers at depth 10, while \emph{T. tinca} was reported as an
XX/XY case with six markers and with no sex-determination system described in prior literature \cite{feron2021radsex}. The same study found no significant
sex-determination signal for \emph{D. albolineatus} or \emph{N. rossii} at depth 10.
Two panels serve as positive controls for male heterogametic
signal recovery; the other two are high-throughput null cases where any
further candidates require cautious ranking.

All four FASTQ-to-table panels come from teleost fishes, and both positive
controls represent male heterogamety, because they are the panels the RADSex
study published, which keeps the inputs fixed across the comparison. The limits that
follow are the method family's rather than this implementation's. RADSex tests
the same two-group presence table, so it too is silent about birds, reptiles,
plants, environmental sex determination, polygenic systems, and
population-specific sex association until someone runs those panels, and it
too needs taxon-appropriate thresholds and biological validation before a call
in a new clade means anything. rsx inherits that boundary; what it changes is
the cost of testing a panel and the amount of evidence returned for one. The
separate amphibian panel below tests downstream marker-table analysis in
another vertebrate class, but it is not a positive control, and female
heterogamety appears here as a modelled direction and a deposited
putative-marker set rather than an experimentally validated ZW panel.
\subsubsection{Independent amphibian marker-table panel}
\label{sec:org13d422a}

Jeffries et al. deposited the Stacks v1 outputs, phenotypic sex map, and
putative XY- and ZW-linked FASTA lists for \emph{Rana berlandieri} alongside their
comparative study of sex-chromosome turnover in true frogs
\cite{jeffries2018truefrogs,jeffries2018truefrogsdata}. The archive contains
11 male and nine female samples. Because it contains Stacks intermediate
files rather than demultiplexed FASTQ, this panel tests the downstream
statistical commands and conversion provenance; it does not enter the
FASTQ-processing timing comparison.

The reproducibility workflow converts 334,411 catalog consensus loci to a
RADSex marker table by summing the deposited sstacks haplotype depths for
each catalog locus and sample. It records the archive MD5, SHA256 hashes of
the catalog, sex map, and 20 match files, and the exact sex-label mapping.
The matrix then runs triage, Yates chi-squared, Fisher exact, Bonferroni, and
marker-level Benjamini--Hochberg FDR at four minimum-depth settings
(1, 2, 5 and 10).

None of the four frequentist combinations returned a marker, and no row
crossed the default posterior threshold (Table \ref{tab:rana_validation}).
The Bayes-factor-only set contracted from 1,102 rows at depths 1--2 to 451
at depth 10. At depth 1, this low-grade set contained 51 of the 71 deposited
putative XY markers and six of the 18 putative ZW markers; the overlaps fell
to one and zero at depth 10. The source study focuses on false-positive
screening across species, and its FASTA labels are putative. We therefore
interpret this panel as a conservative transfer stress test: rsx preserves
weak overlap for audit while withholding a sex-linked call.

\begin{table*}[t]
\centering
\scriptsize
\setlength{\tabcolsep}{2pt}
\begin{tabular}{rrrrrrrr}
\toprule
Depth & Retained & Strict & Post. & BF-only & XY & ZW & FDR \\
\midrule
1  & 311,313 & 0 & 0 & 1,102 & 51/71 & 6/18 & 0 \\
2  & 311,313 & 0 & 0 & 1,102 & 51/71 & 6/18 & 0 \\
5  & 228,175 & 0 & 0 & 963   & 15/71 & 2/18 & 0 \\
10 & 138,311 & 0 & 0 & 451   & 1/71  & 0/18 & 0 \\
\bottomrule
\end{tabular}
\caption{Independent \emph{R. berlandieri} marker-table analysis. Post. denotes
posterior calls. FDR calls are
zero for both Fisher exact and Yates chi-squared tests; Bonferroni calls are
also zero. XY and ZW overlaps compare rsx-emitted evidence rows with the
deposited putative candidate FASTAs. BF-only rows are leads, not calls.}
\label{tab:rana_validation}
\end{table*}

\begin{figure}[htbp]
\centering
\includegraphics[width=0.92\linewidth]{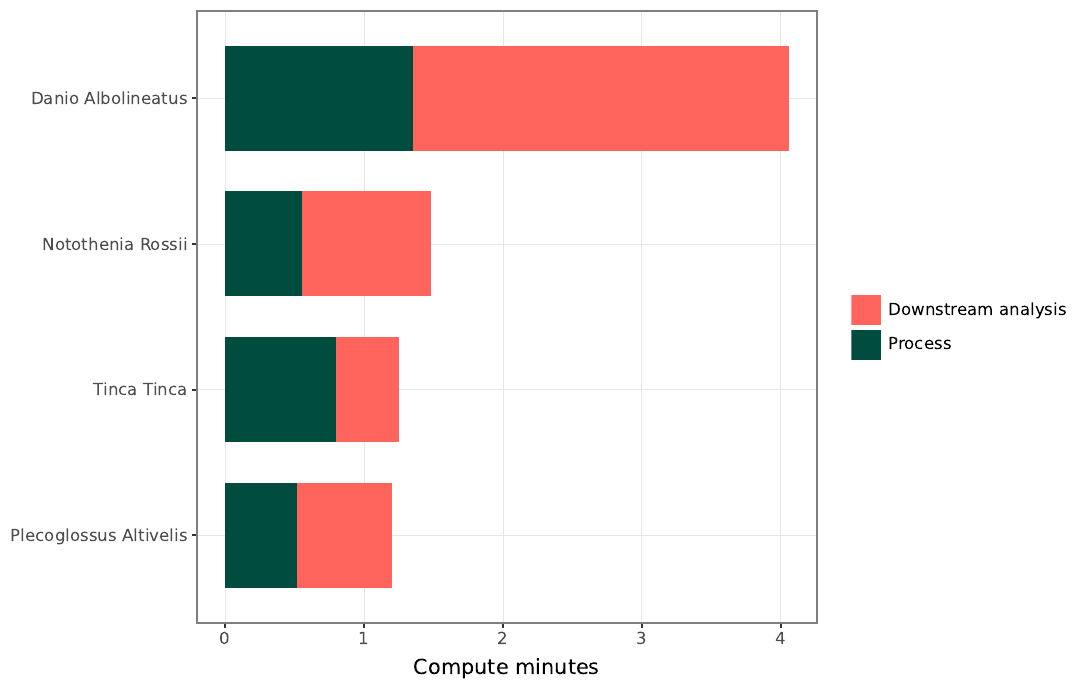}
\caption{Compute-time decomposition for four downloaded literature
RAD-seq datasets. Bars show rsx FASTQ processing and downstream
depth/frequency/distribution/significance analysis at the RADSex
workflow minimum depths.}
\label{fig:lit_compute}
\end{figure}

\begin{figure}[htbp]
\centering
\includegraphics[width=0.92\linewidth]{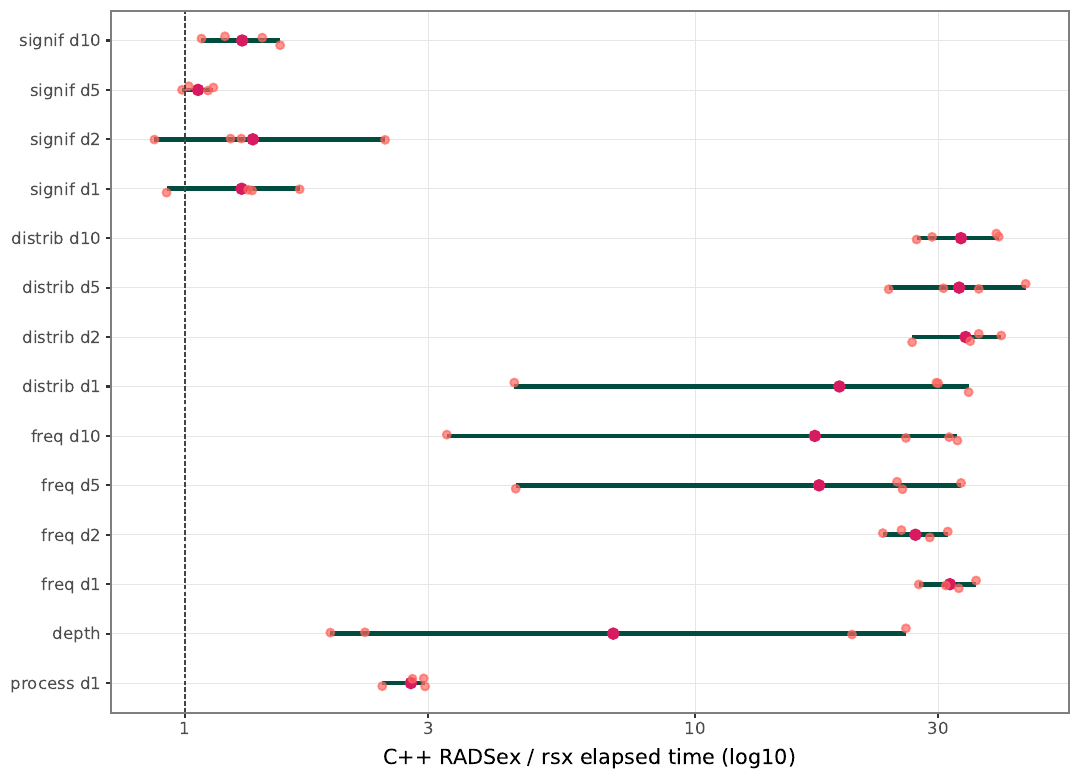}
\caption{Same-input C++ RADSex vs rsx speedups on downloaded literature
datasets. Points are per-dataset ratios; ranges span the minimum and
maximum ratio for each command/depth group, and central points show
geometric means. Values above 1 indicate rsx is faster.}
\label{fig:lit_speedups}
\end{figure}

We ran the identical commands on the downloaded FASTQ files and on the
marker tables produced from them, on a single AMD Ryzen Threadripper PRO
3955WX workstation (16 threads); the deposited reproducibility archive
regenerates these timings, which scale with the host hardware. rsx
ran faster on 53 of 56 command-by-dataset-by-depth pairs. With
speedup defined as C++ wall time over Rust wall time, the geometric mean
is 8.38-fold across the 56 pairs. The process stage alone reaches 2.77-fold
(2.44-fold to 2.96-fold across the four datasets). Table-summary commands
improve the most: 6.91-fold geometric mean for depth, 22.5-fold for frequency, 29.0-fold for
distribution. Significance extraction is the exception (rsx faster on 13 of
16 pairs, 1.25-fold geometric mean) because both programs stream the complete
table (Figure \ref{fig:lit_speedups}).

\subsection{Bayesian marker evidence on literature datasets}
\label{sec:orge27660d}

Alongside the strict thresholded FASTA lists, the Bayesian outputs
return per-marker probabilities of sex linkage. At depth cutoff 10 the
triage rule (Bonferroni-significant or posterior > 0.9) recovers every
strict marker from the two positive-control datasets and widens the
candidate sets without overturning the source-paper classification.
\emph{P. altivelis} adds two posterior-only markers (male-biased, penetrances
0.08--0.27, Bayes factor (BF) 8--16); all 47 strict markers remain.
\emph{T. tinca} adds five.

Each grade points to a different follow-up decision. \emph{P. altivelis} and
\emph{T. tinca} remain XX/XY positive-control datasets. The \emph{P. altivelis}
result agrees with the independently validated amhr2bY sex-determining
gene, but rsx does not validate any individual RAD tag. \emph{D. albolineatus}
remains Bonferroni-negative and contributes 30 W-linked hypotheses for PCR
or mapping. None passes the strict threshold because the depth-10 table
contains 128,109 markers, giving a corrected level of
\(3.9 \times 10^{-7}\). \emph{N. rossii} remains negative because all 400 rows
with \(BF>10\) fall below the posterior boundary. Strict rows meet the
matched statistical criterion, posterior-supported rows form an
unvalidated candidate set, and Bayes-factor-only rows require further
quality control.

\begin{figure}[htbp]
\centering
\includegraphics[width=0.92\linewidth]{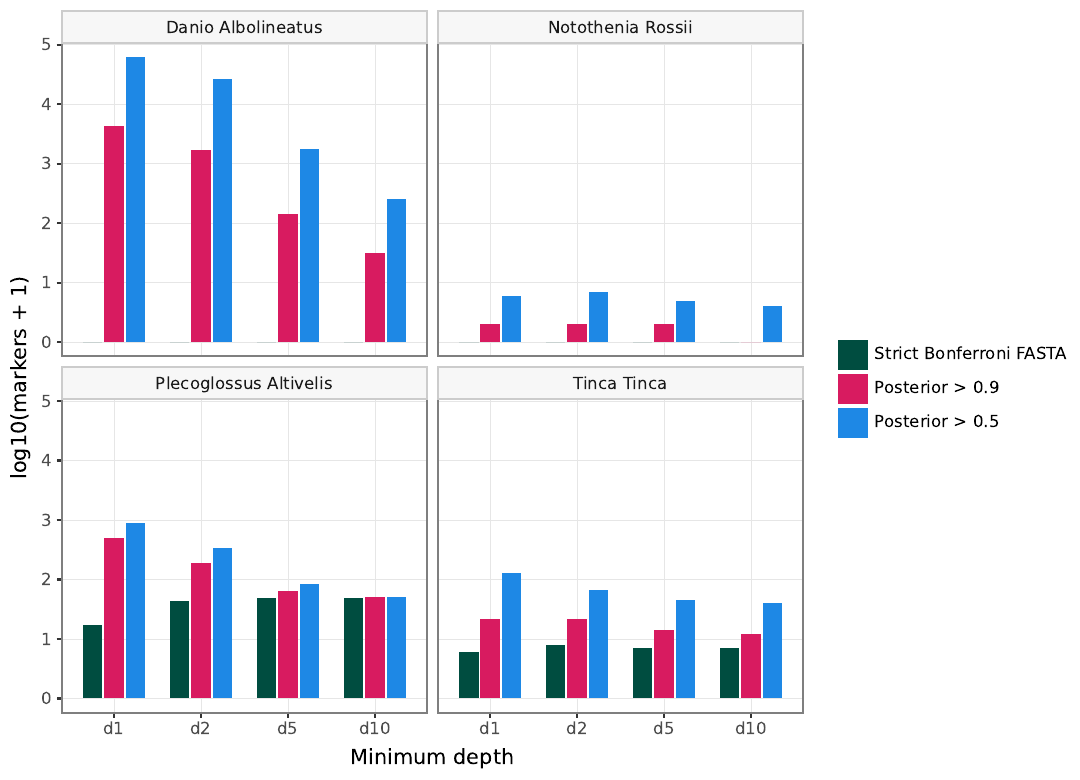}
\caption{Strict and posterior-supported sex-linked marker counts on
downloaded literature datasets.
Strict Bonferroni FASTA extraction is compared with posterior-ranked
candidate counts at the same dataset/depth settings. Each panel is one
published dataset, with minimum depth on the horizontal coordinate and marker counts on
a log10(markers + 1) scale so zero strict calls remain visible.}
\label{fig:lit_candidates}
\end{figure}
\subsection{Alternative inference procedures}
\label{sec:org3090b51}

The procedure chosen for a RAD marker table determines the guarantee attached to each call.

\begin{itemize}
\item Strict Bonferroni \cite{bonferroni1936teoria} plus Yates-corrected \cite{yates1934contingency} chi-squared controls the
family-wise error rate (FWER, the probability of one or more false calls
anywhere in the experiment). A call that a marker is sex-linked therefore
carries a conservative bound on the chance of a false claim across the
whole experiment. Use this path when the result will be presented
as a confirmatory biological fact.

\item Fisher exact \cite{fisher1922interpretation} or G-test \cite{wilks1938large} followed by Benjamini-Hochberg \cite{benjamini1995controlling} false-discovery-rate (FDR) control
trades the stricter family-wise guarantee for higher power. The expected
fraction of false discoveries among all significant markers stays below
the chosen level (5 \% here). This regime fits validation of 20--50
candidates, where a few errors are tolerable if more true
signals are recovered.

\item The Bayesian posterior (the two-component mixture in the Statistical
methods section) does not control a long-run error rate. For each
marker it returns the probability that the marker is sex-linked given
the observed counts and the prior belief \(\pi\) that a random RAD tag
is sex-linked. That probability decides whether the tag is worth the
next PCR or mapping slot.
\end{itemize}

\begin{figure}[htbp]
\centering
\includegraphics[width=0.92\linewidth]{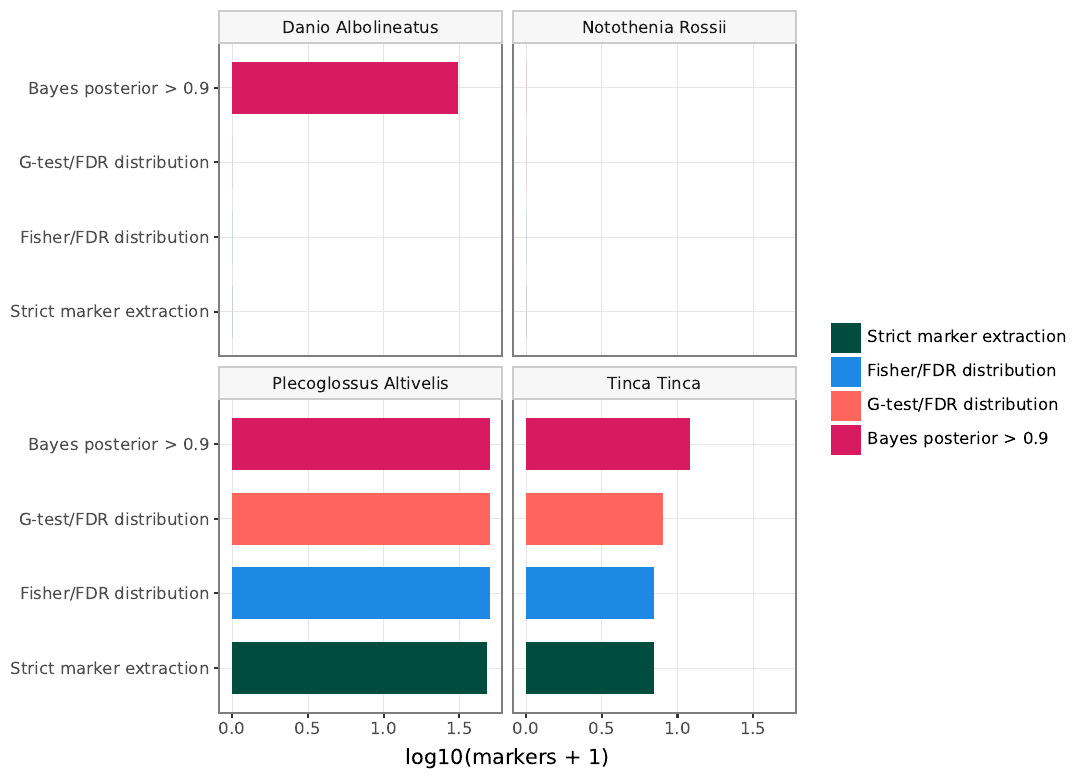}
\caption{Number of sex-linked marker candidates returned by four
different inferential procedures on the identical real marker tables
(minimum depth 10). The vertical scale reports the count of unique RAD tags declared
significant or high-posterior. The four bars per dataset correspond to
(frequentist error-rate control): strict family-wise error rate
(Bonferroni \cite{bonferroni1936teoria} + Yates \cite{yates1934contingency} chi-squared), false-discovery rate (Fisher exact \cite{fisher1922interpretation} +
Benjamini-Hochberg \cite{benjamini1995controlling}), false-discovery rate (G-test \cite{wilks1938large} + Benjamini-Hochberg \cite{benjamini1995controlling}),
and (Bayesian posterior probability): P(sex-linked | data) > 0.9 under
the default mixture model. The figure shows that the choice of procedure
changes the size and composition of the candidate list taken forward for validation, especially in the two panels the
source study treated as negative. /N. rossii/ has no bars because every one
of the four procedures returns zero candidates at depth 10, the expected
result for a panel with no sex-linked signal; its 400 Bayes-factor-only
rows all fall below the posterior $> 0.9$ cutoff and so are reported
separately (Figure \ref{fig:bio_unlocks}).}
\label{fig:mode_candidates}
\end{figure}

On the four real literature tables the three procedures produce visibly
different candidate lists (Figure \ref{fig:mode_candidates}). Fisher/FDR
and G-test/FDR each add markers relative to strict
Bonferroni in the positive-control panels. The Bayesian posterior adds
seven validation candidates in the two confirmatory datasets
while preserving their source-paper XX/XY calls, yields 30
well-supported markers in \emph{D. albolineatus}, a panel the source
publication treated as null, and yields zero high-posterior markers in
\emph{N. rossii}. The triage rule lets one program
serve two scientific goals: strong confirmatory statements
when the data cleanly separate the sexes, and explicit, ranked hypotheses for
validation when the data are marginal. Figure \ref{fig:mode_candidates}
measures how much the choice of inferential procedure changes the
candidate list on real RAD-seq data.

\begin{figure}[htbp]
\centering
\includegraphics[width=0.92\linewidth]{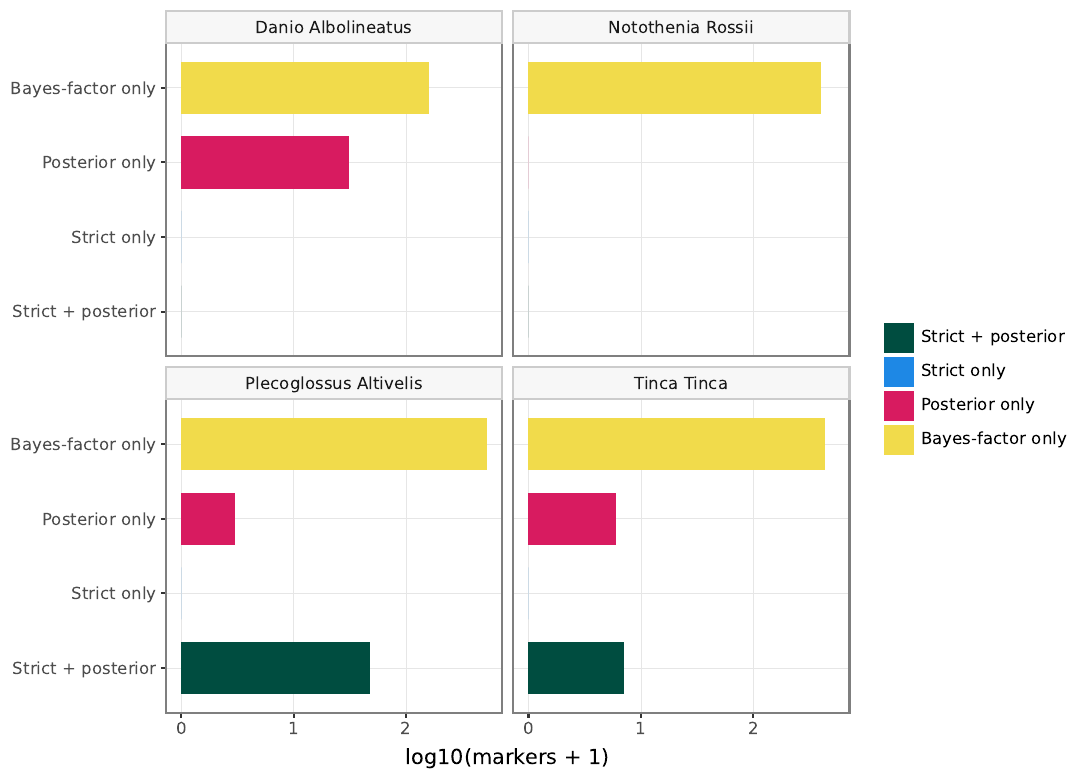}
\caption{Marker-level partition of evidence on the four real literature
panels (minimum depth 10). For each dataset the stacked bars show the
number of distinct RAD tags falling into mutually exclusive categories:
strict (Bonferroni-significant and posterior > 0.9), posterior-only
(posterior > 0.9 but not strict), and Bayes-factor-only (BF > 10 but
posterior $\le$ 0.9). The vertical scale reports the count of unique 80--100 bp RAD tags.
The plot makes visible that the two positive-control datasets are
dominated by overlap between strict and posterior criteria, while
\emph{D. albolineatus} contributes 30 posterior-only rows (marker
hypotheses) and \emph{N. rossii} contributes only Bayes-factor-only rows that
are rejected by the posterior threshold.}
\label{fig:bio_unlocks}
\end{figure}

Figure \ref{fig:bio_unlocks} measures what the posterior
layer adds. In the two positive-control panels the posterior threshold also recovers
the great majority of strict Bonferroni markers (confirmatory overlap),
while a modest number of further
tags cross the posterior > 0.9 line without reaching Bonferroni
significance. In \emph{D. albolineatus} the entire strict category is empty,
yet 30 markers satisfy the posterior criterion; 29 of them are
female-biased, directly supporting the biological claim of a putative
W-linked marker set. Conversely, all 400 BF-only markers in \emph{N. rossii}
fail the posterior threshold. The 52.8 percent singleton fraction provides an
advisory QC measurement and does not filter markers, which limits over-interpretation of a dataset that the source
publication already treated as negative. Figure \ref{fig:bio_unlocks}
partitions the dataset-level inference table (Table
\ref{tab:sex_system_inference}) into the marker classes that drive
each biological call.

The generated inference table included in the reproducibility archive (and
mirrored under the rsx-rs tree) as
\path{results/literature_sex_system_inference.csv} (or the
corresponding file under \texttt{benchmarks/results/}) converts the per-marker
evidence into the four dataset-level biological calls already stated:
confirmatory XX/XY for the two positive controls (with the expected
male-heterogametic direction), a putative W-linked marker hypothesis for
\emph{D. albolineatus}, and no sex-linked marker call for \emph{N. rossii}.

\begin{table}[h]
\centering
\footnotesize
\scriptsize
\begin{tabular}{@{}>{\raggedright\arraybackslash}p{0.19\linewidth}>{\raggedright\arraybackslash}p{0.14\linewidth}>{\raggedright\arraybackslash}p{0.12\linewidth}>{\raggedright\arraybackslash}p{0.45\linewidth}@{}}
\toprule
Dataset & Evidence class & rsx inference & Marker basis \\
\midrule
\emph{P. altivelis} & Positive control & XX/XY-compatible & 47 M-biased strict; 49 M-biased posterior \\
\emph{T. tinca} & Positive control & XX/XY-compatible & six M-biased strict; 11 posterior for validation \\
\emph{D. albolineatus} & Hypothesis & putative W-linked & 30 posterior; 29 F-biased; no Bonferroni call \\
\emph{N. rossii} & Negative & no call & 400 BF-only; zero posterior; 52.8\% singletons \\
\bottomrule
\end{tabular}
\caption{Dataset-level biological inference generated from rsx on
downloaded literature marker tables at minimum depth 10.}
\label{tab:sex_system_inference}
\end{table}

\begin{figure}[htbp]
\centering
\includegraphics[width=0.92\linewidth]{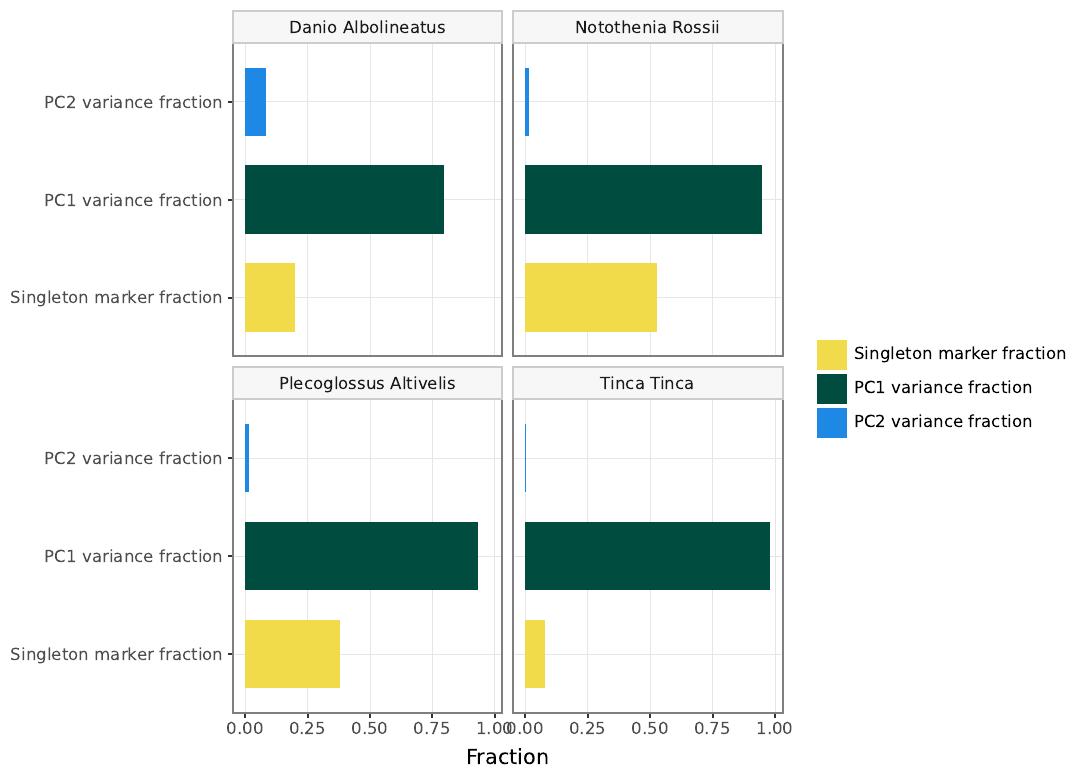}
\caption{Quality-control diagnostics from the rsx frequency and streaming-PCA
modes on the four real literature marker tables at minimum depth 10.
Each panel shows one dataset. Left vertical scale (bars): fraction of markers
observed in only a single individual after minimum-depth filtering
(lower is better for follow-up reliability). Right vertical scale / secondary
elements (lines or points): fraction of total depth variance explained by
the first two principal components of the per-sample depth matrix
(higher PC1 indicates that most variation is library-depth driven rather
than sex-linked signal). See text for interpretation of singleton
fraction and PC loadings as practical QC filters rather than primary
sex-marker detectors.}
\label{fig:mode_qc}
\end{figure}

Figure \ref{fig:mode_qc} reports the quality-control information returned
by the frequency and PCA modes. The singleton fraction spans a wide range:
7.7 percent of the depth-10 markers in \emph{T. tinca} occur in one individual,
whereas 52.8 percent of the \emph{N. rossii} markers are singletons. Candidate
rows from the latter panel therefore require particular scrutiny for
sparse-tag artefacts.

The first two eigenvalue fractions provide a compact scree summary. PC1
accounts for 79.7--98.1 percent of total depth variance across the four
panels, while PC2 contributes much less. The mean male--female coordinate
difference on PC1 is small relative to that concentration of variance.
This combination shows that the leading component does not provide clear
sex separation; it does not, by itself, identify the technical cause.
Users should inspect the per-sample coordinates against library depth and
phenotype metadata, including recorded batch or missingness. The PCA output
contains eigenvalues, variance fractions, and every sample coordinate so
such checks do not depend on the summary plot. PCA supplies a quality-control
and model-diagnostic view separate from the sex-marker tests.

\subsection{Real-data robustness checks on public panels}
\label{sec:org3afb821}

To test whether the graded triage rule and the bounded-memory
implementation hold up beyond the original RADSex workflow panel,
we ran two targeted analyses on real public RAD-seq data, on the same
workstation and with the same processing scripts as the primary
literature results. Both target a place where an evidence-ranking layer
can mislead: dependence on prior settings, and dependence on minimum read
depth.
\subsubsection{Prior sensitivity of the posterior calls}
\label{sec:org556b23c}

We varied the baseline sex-linkage probability over
\(\pi\in\{0.001,0.005,0.01,0.02,0.05,0.1\}\) and the linked-sex detection
probability over
\(p_\text{sex}\in\{0.80,0.85,0.90,0.95\}\). The first range changes prior
odds by about one hundred fold. The second represents marker dropout from
5 to 20 percent in the heterogametic group. These bounds cover screening
designs from rare candidate loci to an unusually enriched RAD panel; they
do not define a universal biological range.

The posterior counts are sensitive to these choices. Across the 24 cells,
the count above 0.9 ranges from 1 to 10,189 for \emph{D. albolineatus}, 0 to 44
for \emph{N. rossii}, 46 to 87 for \emph{P. altivelis}, and 6 to 169 for \emph{T. tinca}.
At the documented default \((\pi,p_\text{sex})=(0.01,0.9)\), the respective
default counts are 30 for \emph{D. albolineatus}; 0 for \emph{N. rossii}; 49 for
\emph{P. altivelis}; and 11 for \emph{T. tinca}. The strict RADSex calls do not use these
parameters and therefore provide the stable reference. Posterior-only
counts, especially for \emph{D. albolineatus}, must be reported with their
chosen prior and treated as hypothesis sets rather than panel
reclassifications.

The 24-cell grid gives a transparent sensitivity report for the present
panels. A study with experimentally validated labels can instead optimize
the parameter vector \((\pi,p_\text{sex},\text{threshold})\) against a prespecified
held-out loss or precision--recall objective. Optuna
\cite{akiba2019optuna} provides seeded samplers, persistent study storage,
and an ask--tell interface that maps each evaluation to an independent
process. On a Slurm system, QCG-PilotJob
\cite{bosak2021qcgpilotjob} can execute those evaluations inside one pilot
allocation, reducing scheduler submissions while preserving one result
record per parameter set. The current four-panel analysis uses the
explicit grid because posterior-only markers lack validation labels for an
optimization objective.

We separately held \(\pi=0.01\), the directional weight at 0.5, and the
posterior boundary at 0.9 while changing the prevalence family. The five TOML
profiles comprised the fixed compatibility rates \((0.9,0.5)\) and a fixed
20-percent-dropout linked rate \((0.8,0.5)\). The first Beta-integrated linked/null
shape pair was \((9,1)/(5,5)\); the other two were \((4,1)/(2,2)\) and
\((1,1)/(1,1)\). These profiles do
not estimate a preferred prior from unlabelled data; they expose how the
candidate set changes when a study encodes different beliefs about detection.

At depth 10, the fixed compatibility profile returned 30 posterior-supported
markers for \emph{D. albolineatus} and none for \emph{N. rossii}; \emph{P. altivelis} yielded
49 and \emph{T. tinca} yielded 11. Under the fixed 20-percent-dropout profile, the
respective counts were 5,403 and 3 for the first two panels, then 57 and 40 for
the latter two. Across the three Beta-integrated profiles, the range was zero
for \emph{D. albolineatus} and 0--2 for \emph{N. rossii}; \emph{P. altivelis} ranged from
52--55 and \emph{T. tinca} returned 7. Thus the positive-control
\emph{P. altivelis} panel remains close to the 49 candidates its default profile
returns, whereas the
\emph{D. albolineatus} hypothesis set is strongly model-dependent. The complete
twenty-row table records every family and parameter rather than presenting a
single posterior profile as biologically privileged.

\begin{figure}[htbp]
\centering
\includegraphics[width=0.98\linewidth]{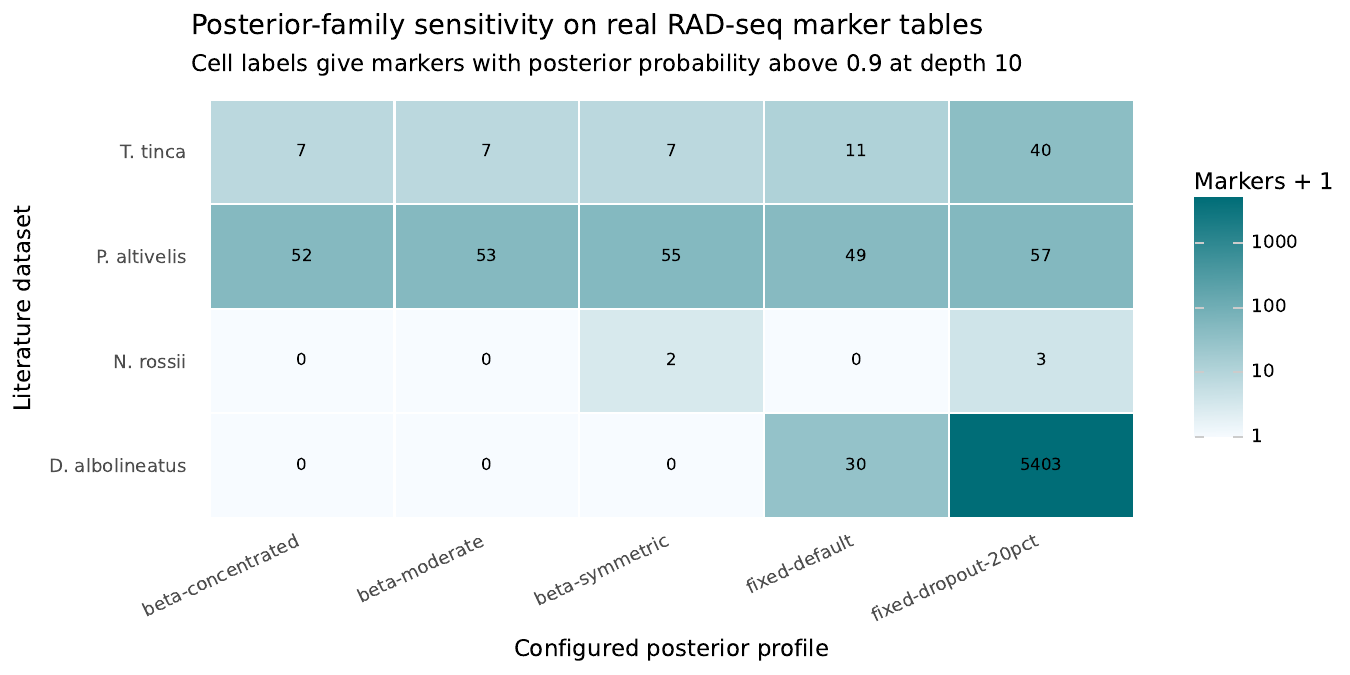}
\caption{Sensitivity to the posterior prevalence family on the four real
literature marker tables at minimum depth 10. Cell labels give the number of
markers above posterior probability 0.9. Every column is a complete TOML
profile with fixed $\pi=0.01$ and equal XY/ZW directional weights; the fixed
and Beta-integrated prevalence parameters are stated in the text and deposited
CSV. These are model-sensitivity results, not optimized biological calls.}
\label{fig:posterior_family_sensitivity}
\end{figure}
\subsubsection{Low-depth behaviour}
\label{sec:org78ba7ed}

A large fraction of published RAD-seq sex-determination studies run
at average depths between 5× and 12× because of cost or
limited input DNA. At these depths the strict Bonferroni threshold
grows more conservative: the number of tests (and with it
the multiple-testing burden) stays large while the per-marker counts
grow noisier. Repeating the full analysis on the real tables at minimum
depths \(\{3,5,8,10\}\) answers a practical question: down to what
sequencing depth does the posterior layer still surface
biologically credible candidates that the strict procedure has already
lost, and where does even the posterior procedure become
unreliable? These curves give an empirical planning guide for
experiments. In the two
positive controls, the strict calls hold at the RADSex depth-10
operating point: \emph{P. altivelis} retains 47 strict markers from depth 5
onward, and \emph{T. tinca} changes by only one strict marker between depths 3
and 10. The posterior and Bayes-factor layers respond more
sharply to depth, exposing how many lower-stringency marker hypotheses
remain available for validation as read support drops.

\begin{figure}[htbp]
\centering
\includegraphics[width=0.92\linewidth]{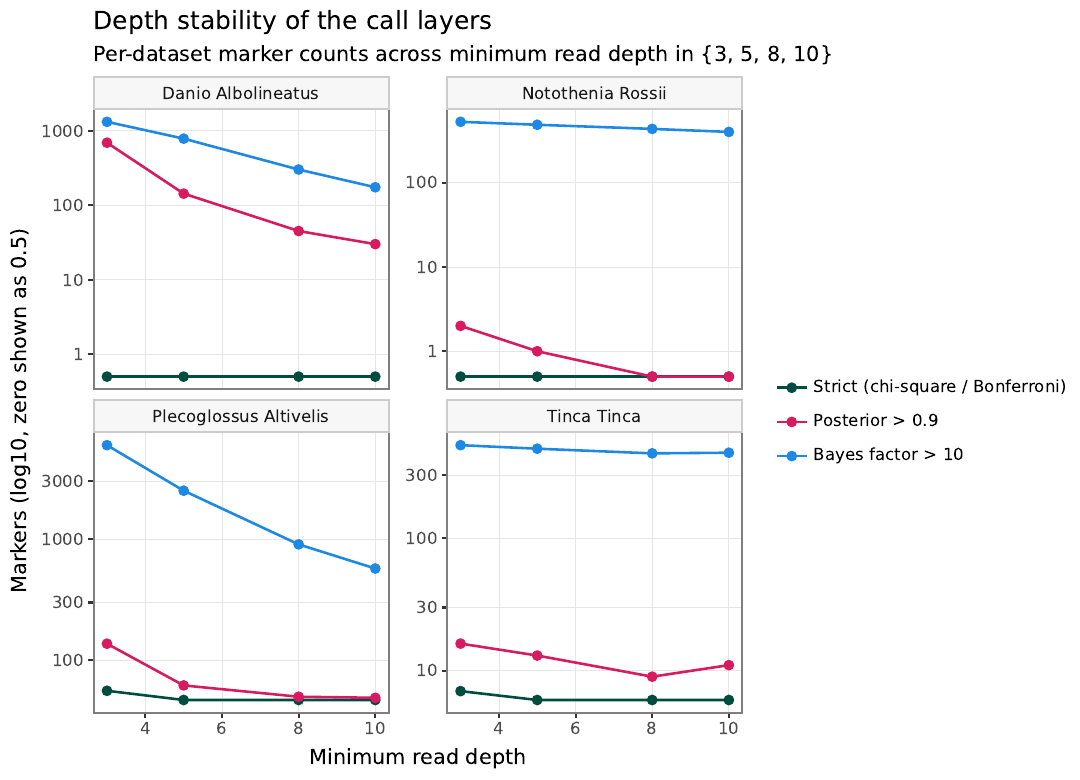}
\caption{How the marker counts respond to the minimum read-depth cutoff, on
the same four literature tables. Raising the cutoff from 3 to 10 discards
tags with too little read support, so every count falls; the question is
whether the conservative, confirmatory calls (strict Bonferroni) hold
steady while the looser candidate sets (posterior $> 0.9$, and Bayes factor
$> 10$) shrink. A stable confirmatory line means the positive-control
signal does not depend on the depth choice, whereas the lower-stringency
sets are depth-sensitive and should be read together with the cutoff. The
counts are drawn on a log scale because they span several orders of
magnitude.}
\label{fig:depth_stability}
\end{figure}

The teleost manifest is included in the reproducibility archive as
\path{benchmarks/literature_datasets.tsv} (and vendored under
the rsx-rs \texttt{repro/} tree). The measured results,
speed comparison tables, Bayesian evidence summaries, marker-level evidence
summaries, binding evidence tables, Slurm scripts, and generated paper
figures that appear in this manuscript match the files in the
\texttt{rsx\_bmc\_repro} archive \texttt{results/} tree (with the same layout under the
rsx-rs \texttt{benchmarks/results/}, \texttt{benchmarks/slurm/}, and \texttt{docs/figures/}
trees). The archive \texttt{README.org} documents the exact
validation and replay steps. The amphibian specification, converter,
portable checksums, per-depth result tables, and compact summary are under
\texttt{config/rana\_berlandieri\_validation.json},
\texttt{scripts/stacks\_to\_radsex.py}, and \texttt{results/rana-berlandieri/}.
\subsection{Software validation}
\label{sec:org68e6897}

We validated rsx at three levels: the mathematics behind each optimized
kernel, the numerical agreement of the kernels with reference values, and
agreement with C++ RADSex on real data.

At the first level, executable symbolic derivations establish that each
shortcut computes the intended quantity. They confirm the
chi-squared-to-erfc identity, the sparse-median selection rule, and the
equivalence between the streaming Gram-matrix PCA and a direct
singular-value decomposition of the centered marker table.

At the second level, numerical tests check that the implementations match
those derivations to the precision the analysis needs. The chi-squared
path reproduces standard critical values from \(p = 0.95\) down to
\(p = 0.001\), and its single- and double-precision routines agree to within
\(10^{-6}\) relative error over that range. Further tests cover odd and even
sparse medians, including half-integer results, on all-zero, single-nonzero,
mixed, and dense inputs; group counts
for up to 200 individuals; Bonferroni and Benjamini-Hochberg behaviour;
the Fisher exact test, the G-test, and the Bayes factor and posterior;
convergence of the empirical-Bayes fit \cite{efron2001empirical}; and each command pipeline on
controlled RAD-seq-like fixtures.

At the third level, rsx and C++ RADSex v1.2.0 run on the same FASTQ and
marker-table inputs across the four literature panels. rsx reproduces the
published RADSex calls, and the Python bindings return the same marker
frequencies/depths and distributions/posteriors as the command-line
tool on the \emph{N. rossii} panel, so the same analysis can be driven from
either interface without changing the result.
\section{Discussion}
\label{sec:org791c530}

rsx is a drop-in successor to RADSex for the marker-table
workflow in reduced-representation sex-determination studies.
Whole-genome, chromosome-assembly, and RNA-seq
approaches answer different biological questions; rsx instead accelerates the
RAD tag workflow with bounded memory and reproducible, inspectable results
from Python and C, while keeping the file formats and command meanings
familiar to existing RADSex users. The implementation therefore separates two
claims. First, a user runs the established RADSex commands with the
same marker-table semantics. Second, the same run returns posterior
evidence grades, streamed quality-control summaries, and binding-friendly
tables that support downstream validation rather than a single
frequentist list.

The performance results follow from this design. Parsing, counting,
depth summarisation, and external sorting benefit most from packed DNA
keys, parallel ingestion, memory-mapped tables, and bounded buffers.
Commands that must score every marker for a global statistical
comparison stay closer to parity, yet they still avoid holding the full
table in memory, except on the FDR ranking path, where global ordering is
intrinsic to the requested calculation. Across the four literature panels,
rsx processed 41.9 billion sequenced bases and marker tables with up to
29 million markers, giving an 8.38-fold geometric-mean speedup across 56
paired RADSex timings and a 2.77-fold speedup for primary FASTQ processing.
These gains are practical, not cosmetic: they cut the cost of
repeating marker discovery across depth thresholds, prior
settings, and validation modes.

The biological results preserve a sharp distinction between matched
statistical calls and unvalidated candidates. rsx reproduces the published RADSex
classifications on the positive-control panels, including recovery of
every Bonferroni-significant marker reported by RADSex at the matched
depth threshold. The posterior layer then adds information about the same
counts without overwriting the frequentist result. Across \emph{P. altivelis}
and \emph{T. tinca} it adds seven posterior-supported candidates to the strict
calls. In \emph{D. albolineatus} it surfaces 30 W-linked hypotheses
from a panel treated as null in the source publication. In \emph{N. rossii} it
withholds 400 BF-only rows from the sex-system inference because their
posterior support remains compatible with a low-prevalence null. This
graded output is the central statistical difference from RADSex: it tells
the reader which rows meet the strict criterion, which become validation
targets, and which carry only marginal evidence.

Those distinctions make the biological panels more informative than a
runtime benchmark alone. The ayu and tench panels test whether rsx can
recover known positive signals while adding only a bounded validation
set. The zebra danio panel tests whether a frequentist null panel still
contains W-linked hypotheses worth targeted follow-up. The
Antarctic notothenioid panel tests the opposite failure mode: many rows
can have large Bayes factors under sparse counts, yet the posterior grade
keeps them out of the sex-system call. The same marker tables therefore
support three uses of rsx beyond speed: reproducing established
statistical calls, triaging weak candidate systems, and guarding against
over-interpretation of rare-tag artefacts.

The \emph{R. berlandieri} reanalysis supplies an independent taxonomic and
file-format transfer test. The Stacks-derived table produces no strict,
FDR-controlled, or posterior-supported call, even though hundreds of rows
have Bayes factors above 10 and some overlap the deposited putative XY and
ZW lists. This separation between evidence grades means that overlap
with an exploratory list does not promote a row to a statistical call.

The sensitivity grid shows where Bayesian rankings change with
\(\pi\) and \(p_\text{sex}\). The positive-control panels retain their strict
RADSex calls because those calls do not use either parameter, whereas the
posterior-only count for \emph{D. albolineatus} changes markedly across the
grid. The depth analysis shows that the strict calls in
the two positive controls remain stable at the usual depth-10 operating
point, while the posterior and Bayes-factor layers expose how validation
candidates expand or contract as read support changes. The QC summaries
and streamed PCA are complementary: singleton fractions and sample-depth
loadings help identify marker tables where rare tags or library-depth
effects dominate before the sex-linked evidence is interpreted.

The numerical implementation supports these biological claims by making
each approximation auditable. The Yates chi-squared path is reduced to an
erfc evaluation for one degree of freedom, and the precision suite measures the
evaluator against the reference rather than adopting its advertised accuracy;
Beta-Binomial and Fisher calculations use log-space forms to avoid overflow. These details are not mathematical
ornament. They form part of the numerical record for a tool
whose output may select markers for PCR, mapping, or breeding decisions.
The reproducibility archive continues a line of literate, reproducible
computational pipelines we have built for biological and high-throughput
analysis \cite{goswamiWailordParsersReproducibility2022,goswamiReproPhylogenetics2023,goswamiHPCNixRedundancy2023}.

Two interpretation limits define how to read the output. FDR correction materializes row data
because global ranking is intrinsic to that procedure; users who need
strictly bounded memory should use Bonferroni or uncorrected streaming
modes. The Bayesian model reports evidence under explicit priors and a
fixed heterogametic-sex prevalence parameter, so posterior-supported
markers remain hypotheses until checked by mapping, PCR, or independent
genomic evidence. rsx makes that boundary visible by separating strict,
posterior-supported, and Bayes-factor-only rows instead of collapsing
them into one marker list.
\section{Conclusions}
\label{sec:orga9939fe}

rsx provides a backward-compatible and statistically extended RADSex
workflow for large RAD-seq sex-determination studies. It keeps the
established marker-table interface and adds bounded-memory ingestion and
sorting, streamed QC and PCA, posterior XY/ZW evidence, Python and C
interfaces, and auditable numerical kernels. On four public FASTQ panels
it reproduces the published RADSex biological classifications,
recovers every matched positive-control marker, processes up to 29
million markers per dataset, and gives an 8.38-fold geometric-mean speedup
across 56 paired timings. On an independently deposited amphibian
marker-table panel it withholds all strict, FDR-controlled, and posterior
calls while retaining low-grade overlaps for audit. The graded output separates strict
Bonferroni calls from posterior-supported validation candidates and
Bayes-factor-only rows. These outputs support use of rsx both as a faster
RADSex replacement and as a marker-prioritisation tool for downstream
experimental validation. The release includes the source code, Python
bindings, C API, derivation scripts, benchmark workflow, full result
tables and figures plus archive checksums needed to reproduce the results
reported here.
\section{Availability and requirements}
\label{sec:orgf15b511}

\begin{itemize}
\item \textbf{Project name}: rsx
\item \textbf{Project home page}: \url{https://github.com/HaoZeke/rsx-rs}
\item \textbf{Archived version}: GitHub releases (\url{https://github.com/HaoZeke/rsx-rs/releases}); the benchmark reproducibility archive used for all results and figures in this paper is deposited on Zenodo under the version-independent DOI \url{https://doi.org/10.5281/zenodo.20531538}.
\item \textbf{Operating systems}: Linux, macOS (Intel + ARM), Windows (without map)
\item \textbf{Programming language}: Rust (>= 1.85); Python bindings require Python
>= 3.9
\item \textbf{Other requirements}: minimap2-enabled builds are required for the
\texttt{map} command; optional MPI and Parquet support are controlled by
feature flags
\item \textbf{License}: GPL-3.0-or-later
\item \textbf{Any restrictions to use by non-academics}: none beyond the GPL license
\end{itemize}
\section{List of abbreviations}
\label{sec:org18503e7}

ABI: application binary interface; API: application programming interface;
BF: Bayes factor; BH: Benjamini-Hochberg; CLI: command-line interface;
CPU: central processing unit; CUDA: Compute Unified Device Architecture;
FASTQ: FASTA with quality scores; FDR: false-discovery rate; FWER: family-wise
error rate; GPU: graphics processing unit; MPI: Message Passing Interface;
PCA: principal component analysis; QC: quality control; RAD-seq: restriction
site-associated DNA sequencing; RSS: resident set size; SBOM: software bill of
materials; SVD: singular value decomposition; TOML: Tom's Obvious Minimal
Language; XY: male heterogamety; ZW: female heterogamety.
\section{Declarations}
\label{sec:org2c7a505}

\subsection{Ethics approval and consent to participate}
\label{sec:orgcd51c80}

Ethics declaration: not applicable. Consent to Participate declaration: not applicable. This computational study collected no new human or animal samples and reanalysed only public, previously deposited sequencing data, so no new ethics approval or consent was required.
\subsection{Consent for publication}
\label{sec:org9302c1d}

Consent to Publish declaration: not applicable. This manuscript does not contain data, images, or details from any individual person.
\subsection{Availability of data and materials}
\label{sec:org477c455}

The rsx source, comprising the Rust core with its CLI and Python bindings, the
tests and the benchmark generation scripts, is archived and openly available
\cite{rsxrs2026}; version 0.2.8 produced every result reported here, and its
\texttt{CITATION.cff} carries the software citation. Version 0.2.9 refits an erfc
evaluator that no code path called and changes no output. A source archive of
the released version accompanies this submission as a supplementary file.

The canonical reproducibility archive for the paper is deposited on Zenodo
\cite{rsxrepro2026}. Its development source tree is the \texttt{rsx\_bmc\_repro} entry
under the reproducibility collection. The deposit contains:
\begin{itemize}
\item the complete \texttt{workflow/Snakefile}, \texttt{profiles/}, \texttt{config/}, and vendored
scripts at the exact revisions used for the manuscript;
\item \texttt{results/} with all CSV files, TSV summaries, and the \texttt{literature\_*.pdf} /
\texttt{.svg} figures that appear in the Results section;
\item the \emph{R. berlandieri} Stacks-to-marker-table converter, archive checksum,
file-level provenance, per-depth result tables, and generated validation
summary;
\item \texttt{rsx\_bmc\_literature\_dataset\_20260604.tar.zst} with the full
literature benchmark run tree, including downloaded FASTQ subsets,
regenerated marker-table workdirs, per-dataset logs, comparison outputs,
binding outputs, and final result tables for the four literature panels
(41.9 billion bases total);
\item the literature dataset manifest and SHA256 checksum sidecars for the
uploaded archives;
\item \texttt{README.org} with the precise commands to inflate the tarball and
replay the full benchmark matrix as a Snakemake \cite{molder2021snakemake}
workflow (\texttt{pixi run snakemake -{}-{}profile profiles/<builder> ...} then the
full target) on any machine with pixi and the matching rsx binary.
\end{itemize}

Lightweight forwarding instructions and older figshare-oriented notes
remain in \texttt{repro/benchmarks.org} and \texttt{repro/literature\_benchmarks.org}
inside the rsx-rs repository; they now point to the \texttt{rsx\_bmc\_repro}
archive as the primary source for all numbers and figures in this paper.

The paper source, an Org-mode single source of truth with its export scripts,
builds reproducibly through \texttt{pixi run mkpdf} for the journal version and
\texttt{pixi run mkarxiv} for the preprint tarball.
\subsection{Competing interests}
\label{sec:org3e2aec2}

The authors declare that they have no competing interests.
\subsection{Funding}
\label{sec:org0cf6f8a}

No specific grant funding was received for this work. The study used institutional research infrastructure at the École Polytechnique Fédérale de Lausanne (EPFL); the institution had no role in study design, data handling, or manuscript preparation.
\subsection{Authors' contributions}
\label{sec:org4f0a9f0}

RG conceived the project, implemented the rsx toolkit and all statistical/Bayesian components, executed the benchmarks on the literature datasets, generated the figures, and drafted the manuscript. RuG contributed the biological interpretation of the marker triage outcomes, cross-checked results against the original RADSex findings, and revised the manuscript for clarity and accuracy. Both authors read and approved the final manuscript.
\subsection{Acknowledgements}
\label{sec:orged31380}

The authors thank the RADSex development team for releasing their C++ implementation and the associated test datasets under an open license. Publicly available RAD-seq data from the cited literature studies are gratefully acknowledged. We also acknowledge EPFL research infrastructure and the open-source scientific computing tools used in the study. This work is dedicated to Ari, Crystee, Jude, Yoda, the garden cats and plants, and family whose support sustains scientific endeavour and efficient implementations.

\appendix
\section{Algorithmic invariants and numerical analysis}
\label{sec:org357f96d}

This appendix records the algorithms and numerical error analyses that
support the performance and correctness claims in the main text. The
material is written at the level needed to verify that the optimisations
preserve the published statistical semantics.

\textbf{\textbf{A.1 Bounded-memory data flow}}

After the initial parallel ingestion of FASTQ files, every subsequent
command works either on a memory-mapped marker table or on external sorted
chunks on disk. The memory footprint is therefore bounded by the number of
phenotypic individuals or by an explicit user-chosen buffer size, never by
the total number of distinct RAD tags. This bounded-memory data flow lets rsx
analyse tables containing millions of markers without down-sampling rare
sex-linked candidates.

The four core patterns are:
\begin{itemize}
\item Streaming producer-consumer pipelines (crossbeam channels between a parser thread and per-marker consumers).
\item Two-pass memory-mapped scans that first count markers for Bonferroni correction, then re-map the same file (kernel page cache) and apply bitset-masked group counts.
\item External sorting (buffered, lz4-compressed runs + k-way merge) for order statistics and table merging.
\item Streaming accumulation of the \(n_{\mathrm{ind}} \times n_{\mathrm{ind}}\) Gram matrix for sample PCA.
\end{itemize}

\textbf{\textbf{A.2 Hybrid strict and Bayesian marker triage}}

Algorithm \ref{alg:triage} is the decision core. The production implementation also:
\begin{itemize}
\item Records the exact (uncorrected) p-value, Bayes factor, and posterior for every emitted row.
\item Writes machine-readable provenance headers containing the exact parameters and the number of markers used for Bonferroni correction.
\item Retains one highest-posterior \texttt{exploratory} row when no marker meets any
configured threshold. This bounded-memory fallback is a ranked lead, not a
sex-linked marker call.
\end{itemize}

Algorithm \ref{alg:triage} is sufficient to reproduce the biological
classifications reported in the results.

\textbf{\textbf{A.3 Two-pass streaming Bonferroni with GroupMask bitsets}}

Algorithm \ref{alg:twopass} gives the full loop. The first pass counts the
number of marker records in the memory-mapped table without materialising
the sequence strings, which fixes the Bonferroni denominator
\(n_{\mathrm{markers}}\) and the threshold \(\tau/n_{\mathrm{markers}}\). The
second pass remaps the same table and evaluates each marker. For marker
\(m\), the implementation intersects the marker's presence bitset with the
two sex-group masks, computes \(g_1\) and \(g_2\) by population count, evaluates
the Yates-corrected chi-squared p-value, and emits the marker only when
\(p < \tau/n_{\mathrm{markers}}\).

The GroupMask bitsets are built once from the popmap (one 64-bit word per
64 individuals) and then reused for every marker. Reuse reduces the
per-marker work from \(O(n_{\mathrm{ind}})\) to \(O(n_{\mathrm{ind}}/64)\) word operations.

Figure \ref{fig:bitset_twopass} states the corresponding \(O(n_{\mathrm{ind}})\)
memory bound.

\textbf{\textbf{A.4 Configurable Bayesian evidence}}

A marker gives $x$ presences among the $m$ individuals of group~1 and $y$ among the $f$ individuals of group~2.

\textit{Bayes factor.} Under $H_1$ the groups have independent rates with priors $p_1\sim\mathrm{Beta}(a_1,b_1)$ and $p_2\sim\mathrm{Beta}(a_2,b_2)$. Under $H_0$ they share $p_0\sim\mathrm{Beta}(a_0,b_0)$. For labelled Bernoulli observations, integrating $k$ presences among $N$ individuals gives
\[ M_{a,b}(k,N)=\int_0^1 p^k(1-p)^{N-k}\frac{p^{a-1}(1-p)^{b-1}}{B(a,b)}\,dp
=\frac{B(k+a,N-k+b)}{B(a,b)}. \]
This gives
\[ \mathrm{BF}_{10}=\frac{M_{a_1,b_1}(x,m)M_{a_2,b_2}(y,f)}{M_{a_0,b_0}(x+y,m+f)}. \]
The compatibility profile uses $(a_j,b_j)=(1,1)$ for all three rates, which recovers the original uniform-prior calculation. All six shapes are runtime parameters and must be finite and positive.

\textit{Posterior probability of sex linkage.} Define $N=m+f$ and the directional concordance counts
\[ k_1=x+(f-y),\qquad k_2=(m-x)+y,\qquad k_0=x+y. \]
$k_1$ counts presences in group~1 plus absences in group~2; $k_2$ reverses that direction; $k_0$ is the pooled presence count for the null. Each hypothesis uses a prevalence family
\[ E(k,N\mid\mathcal P)=
\begin{cases}
q^k(1-q)^{N-k}, & \mathcal P=\mathrm{Fixed}(q),\\[2pt]
M_{a,b}(k,N), & \mathcal P=\mathrm{Beta}(a,b).
\end{cases} \]
For group~1 directional weight $w$, linked evidence and null evidence are
\[ L_{\mathrm{linked}}=wE(k_1,N\mid\mathcal P_L)+(1-w)E(k_2,N\mid\mathcal P_L),\qquad
L_0=E(k_0,N\mid\mathcal P_0). \]
With prior probability $\pi$ that a random tag is sex-linked,
\[ \log O=\log L_{\mathrm{linked}}-\log L_0+\log\frac{\pi}{1-\pi},\qquad
P(H_L\mid x,y)=\frac{1}{1+\exp(-\log O)}. \]
The implementation evaluates the mixture with log-sum-exp and each Beta function through log-gamma. It therefore uses no sampling or numerical quadrature. The compatibility profile sets $\mathcal P_L=\mathrm{Fixed}(0.9)$, $\mathcal P_0=\mathrm{Fixed}(0.5)$, and $w=0.5$; the manuscript benchmarks use that profile. A calibrated study may instead integrate either prevalence with arbitrary positive Beta shapes. The $\lvert\log O\rvert>20$ guard prevents overflow in the logistic transform and does not alter the likelihood model.

\textbf{\textbf{A.5 Streaming Gram PCA via Tucker mode-2 equivalence}}

Let \(X\) be the \(n_{\mathrm{markers}} \times n_{\mathrm{ind}}\) marker-by-sample
matrix, one row \(x_i\) per RAD tag, and let \(\mu\) hold the per-individual mean
depths, so that \(\tilde{X}=X-\mathbf{1}\mu^T\) is the centered matrix.

The usual PCA loadings are the right singular vectors of \(\tilde{X}\), that is,
the eigenvectors of \(\tilde{X}^T \tilde{X}\).

While streaming the table we never form \(\tilde{X}\); we accumulate only the
Gram matrix

\[
  G = \sum_i (x_i-\mu)^T (x_i-\mu) = \tilde{X}^T \tilde{X}.
\]

By the fundamental theorem of SVD (or, equivalently, the mode-2 singular value decomposition of the data tensor viewed in Tucker form), the right singular vectors of \(\tilde{X}\) are exactly the eigenvectors of \(G\).

Derivation (recorded in \path{scripts/sympy/tucker_covariance_proof.py}):

Let \(\tilde{X} = U \Sigma V^T\) be the thin SVD. Then

\(G = \tilde{X}^T \tilde{X} = V \Sigma^2 V^T\)

so the eigenvectors of the Gram matrix are precisely the right singular vectors \(V\) of the centered data matrix. Hence, any eigensolver run on the streamed Gram matrix recovers the identical PCA loadings (up to sign and numerical tolerance) that a full-matrix SVD would have produced.

Memory: \(O(n_{\mathrm{ind}}^2)\) instead of \(O(n_{\mathrm{markers}} \times n_{\mathrm{ind}})\). The same identity is used in the literature for out-of-core PCA on genotype matrices (see also Li et al. 2023 cited in the main text).

\textbf{\textbf{A.6 Forward-error analysis of the Yates chi-squared path}}

For counts \(\le 2^{31}\) the \(2\times 2\) Yates formula is evaluated with
64-bit integer cross-products before any floating-point conversion. The
subsequent floating-point reduction loses at most 1--2 units in the last
place \cite{higham2002accuracy} relative to the exact rational value on the
observed range. The reported CPU results then use the platform erfc
implementation. The precision suite checks that this path tracks erfc to
\(10^{-15}\) and stays within \([0,1]\) across the reported range.
Reported p-values are clamped at \(10^{-16}\) to match the original C++
reference and to avoid spurious ``exact zero'' claims.

The alternative table-driven evaluator is measured the same way. Its 24 panels
carry degree-14 coefficients fitted with Sollya's \texttt{fpminimax}
\cite{chevillard2010sollya}, so the coefficients are already binary64 and the
fitted bound belongs to the numbers rsx carries;
\path{scripts/sollya/erfc_panels.sollya} regenerates them and prints
that bound per panel. The suite then measures the evaluated error against the
platform function over 120,001 points of \([0,6]\) and asserts that neither
evaluator leaves \([0,1]\). Reporting the fitted bound in place of the evaluated
error is not a conservative simplification: an earlier single degree-40 fit
across the same interval carried a fitted bound of \(8.22\times10^{-17}\) and an
evaluated error of \(3.6\times10^{-2}\), because its Horner intermediates reach
\(6^{40}\), and it returned negative values above \(t=4\).

\textbf{\textbf{A.7 Six-significant-digit ``Cg'' floating-point formatter}}

All tables in the paper and in the tool output are emitted with a
six-significant-digit \%g-style formatter that matches the original C++
radsex exactly. The only subtle point is calculating the decimal
exponent. A naïve floor(log10(x)) can be off by one for powers of ten
because of floating-point rounding. The implementation therefore performs
a small verification/adjustment step (see
\texttt{safe\_float\_exponent} in stats.rs). This adjustment was
validated on \(10^6\) random doubles; the formatter was also cross-checked
against the libc output for the same values.

\textbf{\textbf{A.8 Unit-test and script provenance}}

All the above invariants are exercised by the precision test suite
(\path{rsxcore/tests/test_precision.rs}) and by the
SymPy derivation scripts in \path{scripts/} and the Lean model of
the median selection rule. The exact commands used to verify the chi-squared
identity are recorded in the repository so that any future re-implementation
can reproduce the numerical guarantees.

This appendix, together with the algorithms and bounded-memory data-flow
description in the main text, constitutes an implementation-independent
specification of the rsx core algorithms.

\textbf{\textbf{A.9 Exact median from the sparse depth representation}}

Lemma. Let a depth vector have \(n\) entries, of which \(z\) are zero, and let
\(a_0 \le a_1 \le \cdots \le a_{n-z-1}\) be the sorted nonzero entries. For
a zero-based rank \(0 \le r < n\), the entry \(q(r)\) of the sorted full vector is
\(0\) if \(r < z\), and \(a_{r-z}\) otherwise.

Derivation. Sorting the full vector ascending places the \(z\) zeros first, since
no nonzero depth is smaller, followed by \(a_0,\dots,a_{n-z-1}\) in order. The
entry at zero-based rank \(r\) is thus a zero when \(r < z\) and is \(a_{r-z}\)
when \(r \ge z\). \(\qquad\blacksquare\)

Set \(\ell=\lfloor(n-1)/2\rfloor\) and \(u=\lfloor n/2\rfloor\). The exact
mathematical median is \([q(\ell)+q(u)]/2\). For odd \(n\), \(\ell=u\); for even
\(n\), they are the two adjacent middle ranks. Thus, rsx sorts and stores only
the \(n-z\) nonzero depths while returning the same median as the full sorted
vector, including half-integer results. Only the zeros, tracked by their
count \(z\), are elided. Because a RAD tag is absent in most individuals,
\(z/n\) is close to one, and the external-sort volume falls in proportion.

\textbf{\textbf{A.10 Memory mapping, resident pages, and parser bounds}}

Let the marker-table file contain \(L\) bytes and let the operating-system page
size be \(P\). A read-only mapping reserves an address interval of length \(L\);
it does not copy \(L\) bytes into anonymous memory. If \(I_j(t)\) indicates that
file page \(j\) is resident at time \(t\), the file-backed resident set is

\[
  R_{\mathrm{file}}(t)=P\sum_{j=0}^{\lceil L/P\rceil-1} I_j(t),
  \qquad 0\le R_{\mathrm{file}}(t)\le P\lceil L/P\rceil .
\]

The upper bound measures possible cache residency. rsx allocates no anonymous
buffer of length \(L\). Clean pages may be evicted and read again. Linux
anonymous-memory overcommit therefore does not reserve physical memory for the
whole mapped file. Correctness depends only on the immutable byte interval, not
on which pages remain cached between passes.

The parser represents each line and field by offsets into that interval. If a
line occupies bytes \([b_i,e_i)\), parsing advances monotonically from \(b_i\) to
\(e_i\) and allocates no copy of the sequence field. A consumer retains one
decoded marker record plus group masks containing
\(\lceil n_{\mathrm{ind}}/64\rceil\) words. With a channel capacity \(q\) and
maximum record size \(s_{\max}\), parser-owned live storage is bounded by

\[
  M_{\mathrm{parser}}\le (q+1)s_{\max}
    + 16\left\lceil\frac{n_{\mathrm{ind}}}{64}\right\rceil + O(1),
\]

independent of the number of marker rows. The two-pass Bonferroni path scans
the same offsets twice; cached pages can reduce I/O on the second scan but do
not change this storage bound or the selected markers.

\textbf{\textbf{A.11 Two-bit sequence representation}}

Define \(c(A)=0\) and \(c(C)=1\); set \(c(G)=2\) and \(c(T)=3\). For a sequence
\(b_0,\ldots,b_{k-1}\), the packed payload stores

\[
  B_j=\sum_{r=0}^{3} c(b_{4j+r})\,2^{6-2r},
  \qquad 0\le j<\left\lceil\frac{k}{4}\right\rceil ,
\]

with absent positions in the final byte set to zero. A one-byte header stores
\(k\bmod 4\), so decoding recovers the exact length and the payload occupies
\(1+\lceil k/4\rceil\) bytes. Keys through 188 bases fit in the 48-byte inline
representation; longer keys use the same encoding in heap storage. Equality
and hashing cover only the used prefix, so unused inline bytes cannot affect
marker identity.

The optional canonical k-mer signature uses at most 32 bases in a 64-bit word,

\[
  K(b_0,\ldots,b_{k-1})=
  \sum_{i=0}^{k-1}c(b_i)2^{2(k-1-i)}, \qquad k\le 32,
\]

and compares \(K\) with the encoding of its reverse complement. This signature
is an explicitly optional grouping heuristic; the primary packed marker key is
lossless over the four-symbol DNA alphabet.

\textbf{\textbf{A.12 Scope of the cross-method comparison}}

RADSex constructs marker-by-individual depth tables, summarizes frequency and
depth, tests sex-biased distributions, extracts candidates, and optionally
maps them to a reference. SEX-DETector instead uses RNA-seq segregation in
crosses \cite{muyle2016sexdetector}; findZX uses whole-genome coverage profiles
\cite{sigeman2022findzx}; and other low-depth whole-genome methods call
sex-linked scaffolds or karyotypes
\cite{nursyifa2021joint,caduff2024bayesian}. These programs cannot share one
runtime input without manufacturing measurements absent from the original
assays.

We executed SEX-DETector++ and findZX on the empirical examples supplied by
their authors. SEX-DETector++ revision \texttt{00f7d72} completed its standard-model
tests and a 215-step XY analysis of the processed \emph{Silene latifolia} cross; it
classified 38 contigs as autosomal, six as sex-linked, and 18 as lacking
information. findZX revision \texttt{e926637} completed all 110 Snakemake steps of its
no-synteny mantled-howler-monkey example. That example uses subsets of female
read sets SRR9655168 and SRR9655169, male read sets SRR9655170 and SRR9655171,
and the \texttt{AloPal\_v1} reference subset from assembly \texttt{GCA\_004027835.1}. The run
produced the documented QC reports, coverage and heterozygosity tables, and
plot series. These method-native runs establish that the alternate software
was installed and executed; they do not estimate relative biological accuracy.

\begin{table*}[htbp]
\centering
\small
\setlength{\tabcolsep}{3pt}
\begin{tabular}{p{0.15\linewidth}p{0.20\linewidth}p{0.23\linewidth}p{0.28\linewidth}}
\toprule
Method & Required input & Inferential unit & Relation to rsx \\
\midrule
RADSex & RAD-seq reads or marker table & sex-biased RAD tag & matched command and runtime comparator \\
rsx & RAD-seq reads or marker table & sex-biased RAD tag with graded evidence & method evaluated here \\
SEX-DETector & RNA-seq from a controlled cross & sex-linked transcript or allele & complementary segregation analysis; no matched marker-table input \\
findZX & whole-genome read coverage & chromosome or scaffold with Z/X coverage contrast & complementary genome-scale analysis \\
Genome-based association or coverage methods & reference-aligned whole-genome reads & locus, scaffold, or karyotype & complementary when a suitable reference genome exists \\
\bottomrule
\end{tabular}
\caption{Scope of rsx and related sex-linkage methods. Only RADSex shares
the rsx input representation, command meanings, and marker-level target,
so only that pair permits a controlled runtime comparison on identical
inputs. Method-native example runs establish operability, not relative
biological accuracy.}
\label{tab:method_scope}
\end{table*}

\bibliography{rg_main}
\end{document}